\documentclass[fleqn,usenatbib]{mnras}

\usepackage{mathptmx}
\usepackage[T1]{fontenc}
\usepackage{ae,aecompl}

\usepackage{graphicx}	% Including figure files
\usepackage{amsmath}	% Advanced maths commands
\usepackage{amssymb}	% Extra maths symbols

\usepackage{color}
\usepackage{xspace}

\usepackage{pdflscape}
\usepackage{booktabs}
\usepackage{threeparttable}
\newcommand{\ra}[1]{\renewcommand{\arraystretch}{#1}}

\graphicspath{.}

\newcommand{\llambda}{\ensuremath{\lambda \lambda}}
\newcommand{\SN}{\mathrm{S/N}}

\newcommand{\Ha}{{{\rm H}\ensuremath{\alpha}}}
\newcommand{\Hb}{{{\rm H}\ensuremath{\beta}}}
\newcommand{\OII}{{\rm [OII]}}
\newcommand{\OIII}{{\rm [OIII]}}
\newcommand{\NII}{{\rm [NII]}}
\newcommand{\SII}{{\rm [SII]}}

\newcommand{\dex}{\,{\rm dex}}

\newcommand{\Msun}{\,{\rm M}_{\odot}}
\newcommand{\peryr}{\,{\rm yr}^{-1}}
\newcommand{\perGyr}{\,{\rm Gyr}^{-1}}
\newcommand{\Msunyr}{\ensuremath{\,\Msun\, {\rm yr}^{-1}}}
\newcommand{\ang}{\ensuremath{\textrm{\AA}}\xspace}

\newcommand{\Mstar}{{\rm{M}\ensuremath{_{*}}}}
\newcommand{\SFR}{{\rm SFR}}
\newcommand{\sSFR}{{\rm sSFR}}

\newcommand{\flux}{\, {\rm erg}\, {\rm s}^{-1}\, {\rm cm}^{-2}}
\newcommand{\cmcub}{\ensuremath{\,{\rm cm}^{-3} }}

\newcommand{\kmsMpc}{\ensuremath{ \, {  \rm km \, s^{-1} \, Mpc^{-1} }   } }

\title[Evolution of Electron Density with Star Formation Rate]{The COSMOS [OII] Survey: Evolution of Electron Density with Star Formation Rate}

\author[M. Kaasinen et al.]{
Melanie Kaasinen$^{1}$\thanks{E-mail: melanie.kaasinen@anu.edu.au},
Fuyan Bian$^{1,2}$,
Brent Groves$^{1}$,
Lisa Kewley$^{1}$,
Anshu Gupta$^{1}$
\\
$^{1}$Research School of Astronomy and Astrophysics, Australian National University, Weston Creek 2611, Australia \\
$^{2}$Stromlo Fellow\\
}

\date{Accepted XXX. Received YYY; in original form ZZZ}

\pubyear{2016}

\begin{document}
\label{firstpage}
\pagerange{\pageref{firstpage}--\pageref{lastpage}}
\maketitle

% Abstract of the paper no more than 250 words!!!!! Currently 211 words
\begin{abstract}
Star-forming galaxies at $z > 1$ exhibit significantly different properties to local galaxies of equivalent stellar mass. Not only are high-redshift star-forming galaxies characterized by higher star formation rates and gas fractions than their local counterparts, they also appear to host star-forming regions with significantly different physical conditions, including greater electron densities. To understand what physical mechanisms are responsible for the observed evolution of star-forming conditions we have assembled the largest sample of star-forming galaxies at $z\sim 1.5$ with emission-line measurements of the $\OII \llambda 3726,3729$ doublet. By comparing our $z\sim 1.5$ sample to local galaxy samples with equivalent distributions of stellar mass, star formation rate and specific star formation rate we investigate the proposed evolution in electron density and its dependence on global properties. We measure an average electron density of $114_{-27}^{+28}\cmcub$ for our $z\sim 1.5$ sample, a factor of five greater than the typical electron density of local star-forming galaxies. However, we find no offset between the typical electron densities of local and high-redshift galaxies with equivalent star formation rates. Our work indicates that the average electron density of a sample is highly sensitive to the star formation rates, implying that the previously observed evolution is mainly the result of selection effects.
\end{abstract}


%%%%%%%%%%%%%%%%%%%%%%%%%%%%%%%%%%%%%%%%%%%%%%%%%%

%%%%%%%%%%%%%%%%% BODY OF PAPER %%%%%%%%%%%%%%%%%%

\section{Introduction}

	The cosmic star formation rate has changed significantly since the formation of the first galaxies, declining by an order of magnitude in the last 10 Gyr \citep{2014ARA&amp;A..52..415M}. Not only did the early Universe ($z>2$) contain a greater fraction of actively star-forming galaxies, it also hosted star-forming galaxies with considerably higher star formation rates (SFRs) than galaxies of equivalent stellar mass (\Mstar) today \citep{2011A&amp;A...533A.119E,2007ApJ...670..156D,2014ApJS..214...15S}. Although we now know that star-forming galaxies have grown in size, accumulated stellar mass and become more chemically enriched over cosmic time, we are yet to understand what mechanisms have driven the changing star formation rate.

	To unlock the star formation history of the Universe we need to better understand the conditions within star-forming regions. %These conditions are largely governed by the feedback between ionized gas clouds within the interstellar medium (ISM) and the stars which inhabit them. % Stars deposit energy and heavy elements into the ISM. 
	Both the ionizing sources and physical conditions of star-forming regions can be probed by measuring the strengths of rest-frame optical emission lines stemming from the ionized gas.  
	The relative strengths of these emission lines are mainly governed by a small set of ISM properties including %
	the chemical abundance, shape of the ionizing radiation field, ionization state and gas density   %
	\citep{2002ApJS..142...35K,0067-0049-167-2-177,0004-637X-647-1-244,0004-637X-774-2-100}.

	Prior to the last decade, samples of rest-frame optical spectra of $z>1$ galaxies were small, biased and only included the strongest emission lines. The lack of near-infrared (NIR) spectroscopy for high-redshift galaxies reflected the difficulty in combating the combined effects of poor CCD sensitivity, detector fringing, and sky deterioration, resulting in a cosmological blind spot at $1<z<3$ dubbed the ``redshift desert'' \citep{2004ApJ...604..534S,2004ApJ...607..226A}.  Far from being an arid landscape, the redshift desert is an important epoch in the history of star-formation and galaxy assembly in the Universe. Not only did the star formation rate density peak around $z\sim 2$ \citep{2014ARA&amp;A..52..415M,2006ApJ...651..142H} but the majority of the stellar mass and heavy elements in today's Universe were produced in the redshift desert \citep{2003ApJ...587...25D,2005ApJ...619L.131D,2003ApJ...594L...9F,2003ApJ...599..847R,2004ApJ...617..240K,0004-637X-792-1-75}.

	Thanks to the advent of NIR multi-object spectroscopy on 8-10 m telescopes, large samples of $z > 1$ star-forming galaxies with rest-frame optical emission-line measurements are now being assembled. The ensuing studies suggest that high-redshift star-forming galaxies exhibit emission-line properties that are atypical of the local star-forming galaxy population. In particular, there is increasing evidence for a significant enhancement of emission-line ratios including $\OIII 4959,5007/\OII 3726,3729$ and $\OIII 5007/\Hb$ at high-redshift \citep[e.g.][]{0004-637X-701-1-52,0004-637X-774-2-100,2014ApJ...795..165S,2015PASJ...67...80H,0004-637X-801-2-88}. The observed changes in emission-line ratios indicate that at least some of the conditions within star-forming regions must have evolved since the early Universe.

	It is still unclear which physical properties of star-forming regions are driving the observed changes in emission-line ratios. Although high-redshift galaxies are less chemically enriched than local galaxies of the same stellar mass, \citep{0004-637X-792-1-75,2004ApJ...617..240K,2014MNRAS.440.2300C}, the elevated line ratios cannot be solely attributed to lower chemical abundances \citep[e.g.][]{2014MNRAS.442..900N,2014ApJ...795..165S,2014ApJ...785..153M,2009ApJ...701...52H,0004-637X-774-2-100}. Other possible contributors include higher ionization parameters and/or electron densities \citep[e.g.][]{2008MNRAS.385..769B,2015ApJ...812L..20K,2014ApJ...785..153M,2014ApJ...787..120S}, harder ionizing radiation fields \citep{2014ApJ...795..165S,0004-637X-774-2-100}, contributions from shocks/AGN \citep{2006MNRAS.371.1559G,2014ApJ...781...21N}, a variation in N/O ratio \citep[e.g.][]{0004-637X-801-2-88} and/or selection effects \citep{2016arXiv160601259D,2014ApJ...788...88J,2016ApJ...817...57C,2014ApJ...781...21N,0004-637X-774-2-100}. 

	To resolve the current deadlock, we have assembled a sample of star-forming galaxies at high-redshift ($z\sim 1.5$) with the ``full suite'' of rest frame optical line flux measurements (i.e. \OII, \Hb, \OIII, \Ha\ and \NII). Thanks to the FMOS-COSMOS survey \citep{2014arXiv1409.0447S,2016arXiv160406802K}, measurements of \Hb, \OIII, \Ha\ and \NII\ are already available for a large ($\sim 500$) sample of galaxies at $z\sim 1.5$. We complement the existing FMOS data with high resolution ($R\sim 2000$) spectroscopy of the \OII $\lambda 3726,3729$ doublet. Previous studies at $z\sim 1.5$ were either statistically insignificant, plagued by selection effects or lacked the necessary number of emission lines required to probe the conditions within star-forming regions \citep[e.g.][]{2015PASJ...67...80H,2016arXiv160406802K,2013ApJ...777L...8K,2008ApJ...678..758L}. By gathering the largest sample of star-forming galaxies at $z\sim 1.5$ with \OII\ doublet measurements our work provides the missing link between the local Universe and recent high-redshift ($z\sim 2.3$) studies from the KBSS and MOSDEF surveys \citep[e.g.][]{2016ApJ...816...23S,0004-637X-801-2-88}. Here, we present our data and measurements of the electron density.  

	Increasingly, high-redshift observational studies are finding evidence for elevated electron densities and/or ionization parameters \citep[e.g.][]{0004-637X-701-1-52,2010ApJ...725.1877B,2014MNRAS.440.2201S,2014ApJ...787..120S,2016ApJ...816...23S,2014ApJ...785..153M}. However, the current body of work remains inconclusive, with most studies limited by small sample sizes and/or selection effects. Because most high-redshift studies fail to take out correlations with global galaxy properties when comparing the typical electron densities and ionization parameters of high-redshift and local galaxies it remains unclear whether the enhanced electron densities and ionization parameters at high redshift are simply the byproduct of probing ``typical'' star-forming galaxies with higher specific star formation rates (sSFRs, \SFR/\Mstar) than ``typical'' local galaxies. 

	Recently, several studies have suggested a correlation between the electron density and/or ionization parameter of star-forming galaxies and their specific star formation rates \citep{2015ApJ...812L..20K,2016ApJ...820...73H,2016ApJ...822...62B}. We investigate the proposed correlations with global galaxy properties by comparing the electron densities of our $z\sim 1.5$ sample to three samples of local star-forming galaxies with equivalent distributions of either stellar mass, \SFR\ or both mass and SFR. By matching the local and high-z ($z\sim 1.5$) samples based on their global properties we are able to isolate the primary driver of the observed increase in electron density.

	This paper is structured as follows. In section \ref{sec:Obsred} we describe the survey design, observation and data reduction for our $z\sim 1.5$ sample. We describe the selection of the high-z and local samples used in this work in section \ref{sec:SampleSel} and show the global properties of these samples. In section \ref{sec:Eldens} we estimate the typical electron densities of our samples and investigate the proposed evolution with redshift. We conclude by summarising our results in section \ref{sec:Summary}. Throughout this paper we refer to values of \SFR, \sSFR\  and \Mstar\ consistent with a Kroupa IMF. When deriving the \SFR, we adopt a $\mathrm{\Lambda}$-CDM cosmology with $H_0=70\kmsMpc$, $\Omega_m = 0.3$, and $\Omega_\Lambda = 0.7$.

%%%%%%%%%%%%%%%%%%%%%%%%%%%%%%%%%%%%%%%%%%%%

\section{Observations and Data Reduction} 
\label{sec:Obsred}

	\subsection{The COSMOS-[OII] Survey} 

		Our work is based upon a sample of star-forming galaxies from the COSMic evOlution Survey (COSMOS). We derive our sample from COSMOS to take advantage of the extensive multiwavelength ground and space-based observations already at hand \citep{2007ApJS..172....1S}. The 2 square degree equatorial field encompassed by COSMOS is visible from most ground based optical and IR telescopes including the Keck and Subaru telescopes \citep{2007ApJS..172....1S}. A major spectroscopic survey has already been undertaken using the Fibre Multi-Object Spectrograph (FMOS) on Subaru (PIs Sanders and Silverman, \citealt{2014arXiv1409.0447S}), resulting in \Ha\ detections at $\SN>3$ for $\sim 900$ galaxies at $1.4<z<1.7$. We complete the emission-line measurements for these galaxies with corresponding observations of the \OII \llambda 3726,3729 doublet using the DEep Imaging Multi-Object Spectrograph (DEIMOS) \citep{2003SPIE.4841.1657F} on Keck II.

		Our COSMOS [OII] survey (PI L. J. Kewley) is primarily targeted at galaxies with existing FMOS spectroscopy. All targets were identified from the COSMOS photometric catalogues \citep{2012A&amp;A...544A.156M,2013A&amp;A...556A..55I}. Initial stellar masses for target selection were estimated based on the broad-band photometry and fitting results of LePHARE \citep{2011ascl.soft08009A}. We targeted galaxies with $\Mstar \geq 10^{9.8} \Msun$ (Chabrier IMF), $\SFR_\mathrm{phot} > 10\Msunyr $  and z(AB) magnitudes $\lesssim 24$ (SuprimeCam, $z^{++}, \lambda_c = 9106$, \citealt{2016arXiv160402350L}).  For further analysis we use the latest stellar mass estimates from \cite{2016arXiv160402350L}, normalised to a Kroupa IMF.

    \subsection{COSMOS-[OII] Observations} 

    	Spectroscopic observations for the COSMOS \OII\ survey were conducted over two nights, UTC February 24 and 25, 2014, with DEIMOS on the Keck II telescope. We observed seven COSMOS masks, each with $\sim 150$ slits. Each mask was observed three times for 20 minutes, to optimise cosmic ray rejection. The average seeing over the two nights was $\sim 0.75"$. All observations were conducted with the 600ZD grating centred at $7500 \ang$, the OG550 filter and $1"$ slit width. For the $1.4 \lesssim z \lesssim 1.7 $ galaxy sample in this paper, the \OII\ doublet falls on the central portion of the red side (7500-9800\ang) of the DEIMOS spectrograph. The resulting spectra have a dispersion of $\sim 0.65\ang/\mathrm{px}$ and spectral resolution of $R\sim 2000$. 

    \subsection{Data Reduction} % (fold)
        \label{sub:raw_data_reduction}

    	% Paragraph 1: raw data reduction - idl pipeline
    	Our raw science frames were initially processed using the publicly available IDL based pipeline, \emph{spec2d}, developed by the DEEP2 survey team \citep{2012ascl.soft03003C,2013ApJS..208....5N}. The \emph{spec2d} pipeline performs bias removal, flat fielding, cosmic ray rejection, slit-tilt corrections and wavelength calibration on a slit-by-slit basis \citep{2013ApJS..208....5N}. Sky subtraction was performed without the use of a dithering pattern. We used standard Kr, Xe, Ar, and Ne arc lamps for wavelength calibration. Using \emph{spec2d} we generated one sky-subtracted, cosmic-ray cleaned two-dimensional (2-d) spectrum for each slit. 
    	Our initial \OII\ sample was detected by visually inspecting the 2-d spectra for \OII\ emission features. For galaxies where emission lines were present we recorded the central position on the slit of the emission feature (and, where present, the continuum) as well as an initial redshift estimate based on the observed wavelength of the \OII\ doublet. Where relevant, we noted the presence of multiple emission features (from separate star-forming regions) and any data reduction problems such as sky continuum errors caused by scattered OH light from a neighbouring slit \citep[see][]{2013ApJS..208....5N}. Based on the initial redshift estimates and additional notes we selected 115 galaxies for further analysis. 

    	We simultaneously corrected our 2-d spectra for detector sensitivity and atmospheric extinction using observations of the flux standard star, DA white dwarf G191-B2B \citep{1990AJ.....99.1621O}. From the 1-d stellar spectrum we derived a sensitivity curve, representing the convolution of the instrument response function and the atmospheric absorption. All 2-d spectra were divided by the sensitivity curve to remove the effects of telluric absorption and instrument response. We derived a flux scaling relation by applying the sensitivity curve to the 1-d spectrum of G191-B2B and matching the corrected spectrum with the absolute flux spectrum from the ESO archives. The resulting scaling relation was applied to all 2-d slit spectra to generate 2-d flux calibrated spectra.
   
    	The 2-d flux calibrated spectra were reduced to 1-d by calculating the total flux over the effective aperture for the ``red'' side of each spectrum. We define the effective aperture as the region along the length of a slit at which emission-line features or continua are detected (horizontal dashed white lines in the upper panels of Fig. \ref{fig:spec}) and determine the size of the effective apertures by fitting gaussian profiles to the spatial flux distributions. We summed the flux within the effective aperture over the ``red side'' of each spectrum to produce our 1-d spectra (bottom panels Fig. \ref{fig:spec}). Small changes to the size of the effective aperture had no measurable impact upon the derived 1-d galaxy spectra and were therefore not considered in our error calculations. 

    % section extracting_1d_spectra (end)

    \subsection{Emission line fitting} % (fold)
        \label{sub:emission_line_fitting}

    %  OII emission line fitting
    We measure the emission-line fluxes of the \OII\ doublet by fitting a double gaussian profile to a $\sim 35 \ang$ window of each 1-d spectrum, centred at the observed wavelength of the \OII\ doublet (bottom panels Fig. \ref{fig:spec}). Our fitting routine utilises IDL's `MPCURVEFIT', a Levenberg-Marquardt least squares minimization algorithm which fits a user supplied model and returns best-fit parameters, errors, and a measure of the overall quality of the fit. We fix the vacuum wavelengths of the doublet lines to $\lambda_1 = 3727.09\ang $ and $\lambda_2 = 3729.88 \ang$.  To minimise the impact of OH lines our fitting routine takes into account a $1/N^2$ weighting scheme for each pixel, based on the standard deviation of the flux outside the effective aperture ($N$, blue lines in Fig. \ref{fig:spec}). Throughout, we use the 1-sigma error on the fit parameters combined with the covariance values returned by `MPCURVEFIT' to estimate the errors on our parameters. The reduced chi-squared values ($\chi^2/\nu$) returned by our fitting routine are used as an indication of the goodness of the line fits and our errors are scaled accordingly for fits where $\chi^2\nu>1$. 

%%%%%%%%%%%%%%%%%%%%%%%%%%%%%%%%%%%%%%%%%%%%%%%%%%%%%%%%%%

\section{Sample Selection}
	\label{sec:SampleSel}

		\begin{figure*}%[!t]
              \begin{center}  
                \includegraphics[width=0.95\textwidth,trim={0cm 0cm 0cm 0cm},clip]{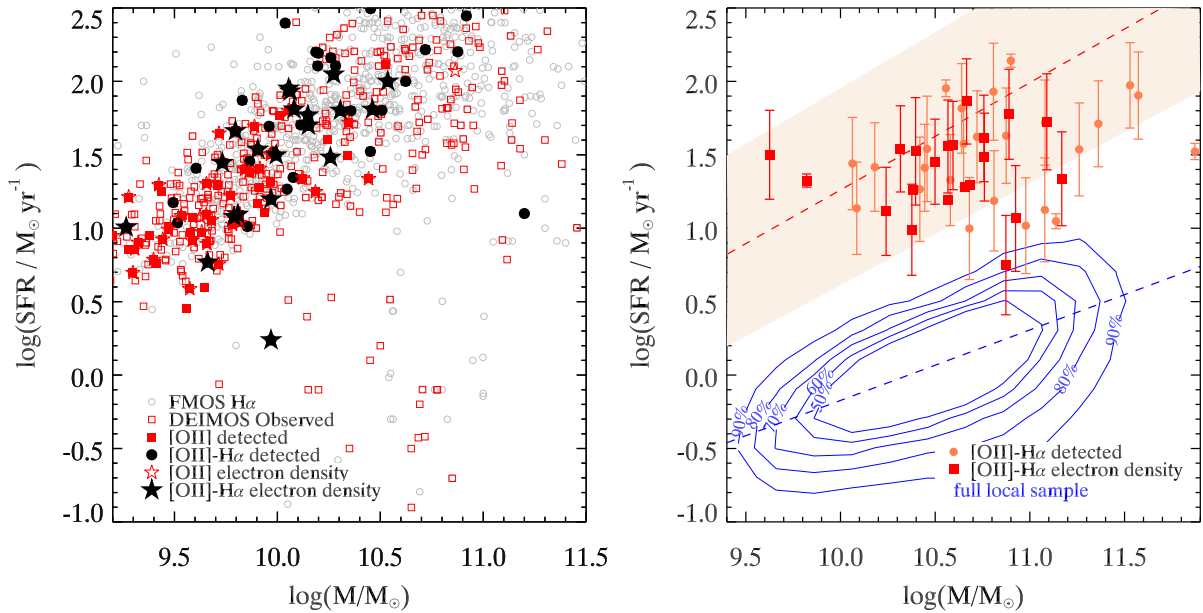} %
                %trim: left lower right upper
                \end{center}
            \caption{ %
            		SFR vs \Mstar\ for the high-z samples discussed in this work. Left panel: median $\log(\SFR)$ and $\log(\Mstar)$ determined by Laigle et al. (2016) based on fits to photometry using LePHARE. Galaxies observed as part of the FMOS-COSMOS (open grey circles) and COSMOS-\OII\ surveys (open red squares) are  compared to \OII-detected (filled red squares) and \OII-\Ha\ detected (filled black circles) subsamples. The samples for which electron densities could be determined are shown as stars. Right panel: \Ha\ based star formation rates for the $z\sim 1.5$ \OII-\Ha\ sample (filled orange circles), the electron density subsample (red squares) and local SDSS star-forming galaxies (blue contours, showing the distribution density). The local and $z\sim 1.5$ samples are compared to the main-sequence fits at $z\sim 1.5$ (red dashed and filled pink) and $z\sim 0$ (blue dashed) given by equation~28 of Speagle et al. (2014). Note that the left and right panels are not directly comparable due to differences between photometric and \Ha\ based SFRs.  
						}  %
            \label{fig:ms} 
        \end{figure*}

	 	\begin{figure*}%[!t]
              \begin{center}  
                \includegraphics[width=0.92\textwidth,trim={0cm 2cm 2cm 0cm},clip]{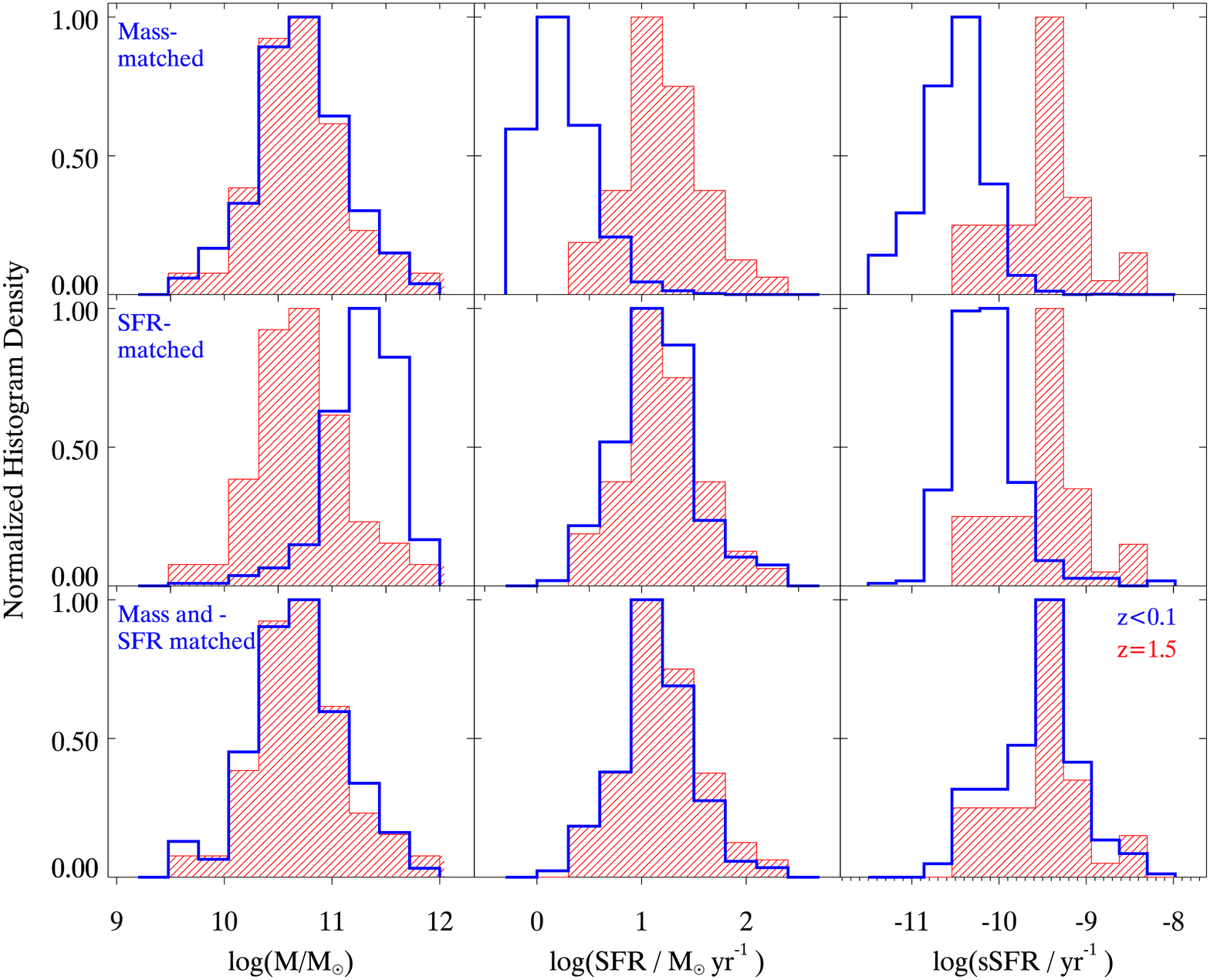} %
                %trim: left lower right upper
                \end{center}
            \caption{Comparison between the distributions of \Mstar\ (left), \SFR\ (centre) and \sSFR\ (right) for the matched local comparison samples (blue outline) and $z\sim 1.5$ \OII-\Ha\ sample (red line fill). Top row: \Mstar-matched local sample. Middle row: SFR-matched local sample. Bottom row: \Mstar-and-SFR-matched local sample.  }  %
            \label{fig:match} 
        \end{figure*}

	\subsection{The $z\sim 1.5$ Sample} 

		Our COSMOS \OII\ survey targeted $\sim 430$ galaxies at $1.4 \lesssim z \lesssim 1.7$, 103 of which were identified to have [OII] detections ($>2\sigma$). Of the 103 galaxies with detected [OII], 46 galaxies have corresponding \Ha\ detections from the FMOS-COSMOS survey. We refer to the total sample of 103 COSMOS galaxies for which we detect \OII\ as the ``\OII\ sample'' and refer to the subsample for which we have detections of both \Ha\ and \OII\ as our ``\OII-\Ha\ sample''. We present the data, derived quantities, global properties and spectra of our \OII-\Ha\ sample in Table \ref{tab:main_sample} and Fig. \ref{fig:spec}.

		We take the median stellar masses from \cite{2016arXiv160402350L}, derived by fitting model spectra to the spectral energy distributions via LePHARE \citep{2011ascl.soft08009A} following the methods outlined by \cite{2015A&amp;A...579A...2I}. To convert to a Kroupa IMF we apply a constant scaling factor of $1.06$, taken from \cite{2012ApJ...757...54Z}. The SFRs and sSFRs of our \OII-\Ha\ sample were estimated from the dust-corrected \Ha\ luminosities using the conversions in \cite{2011ApJ...737...67M} and \cite{2011ApJ...741..124H} (consistent with a Kroupa IMF). Dust corrections were estimated using $\Ha/\Hb$ assuming an instrinsic ratio of 2.86, consistent with Case B recombination at $T=10^4\mathrm{K}$ and $n_e=100\cmcub$ \citep{2003adu..book.....D,2006agna.book.....O}, and the \cite{1989ApJ...345..245C} extinction curve.

		The \OII-\Ha\ sample appears to be representative of the range of SFRs spanned by $z\sim 1.5$ star-forming galaxies at the same stellar masses (see Fig. \ref{fig:ms}). As in previous high-z studies, our \OII-\Ha\ sample exhibits higher SFRs than is typically found for local star-forming galaxies of the same stellar mass (red data vs. blue contours, right panel of Fig. \ref{fig:ms}). The SFRs of our \OII-\Ha\ sample are mostly consistent with relation at $z\sim 1.5$ given by the best main-sequence fit derived by \cite{2014ApJS..214...15S} (see equation 28 therein). We describe the source of the scatter within the \OII-\Ha\ sample in Kaasinen et al. (in prep.). 

		Furthermore, the stellar masses and SFRs of the \OII-\Ha\ sample are consistent with those of the parent FMOS and COSMOS-[OII] samples (left panel, Fig. \ref{fig:ms}). Because many of the galaxies in the COSMOS-\OII\ sample do not have \Ha\ detections we compare the SFRs determined photometrically by \cite{2016arXiv160402350L}. The SFRs in the left and right panels of Fig. \ref{fig:ms} are therefore not comparable (in fact there exists significant scatter in the correlation between the two). The \OII-\Ha\ sample spans the stellar masses range $10^{9.6} - 10^{11.9}\Msun$ and SFR range $3-150\Msunyr$ with a median stellar mass of $10^{10.7}\Msun$ and median SFR of $15\Msunyr$. The range of specific star formation rates spanned by the \OII-\Ha\ sample is $0.04-7.3\perGyr$, with a median sSFR of $0.4\perGyr$.

	% subsection the_z=1.5 sample (end)
				
	\subsection{Local Comparison Samples} % (fold)

		To investigate the evolution in electron density we compare our $z\sim 1.5$ \OII-\Ha\ detected sample to three local comparison samples, matched according to their global properties. Previous high-redshift observational studies \citep[e.g.][]{2016ApJ...816...23S,2011ApJ...732...59R} fail to take into account the evolution of global galaxy properties when drawing comparisons between the ISM conditions in the local and high-redshift Universe. Yet, the local star-forming galaxy population typically has far lower SFR and sSFR than the high-z population to which they are compared. Because most previous studies compare significantly different populations of galaxies, it remains unclear to what extent the global attributes of galaxy samples are responsible for driving the observed evolution of ISM conditions. We address this issue by selecting three local comparison samples matched to our primary high-z sample according to the global properties commonly used to design galaxy surveys (i.e. \Mstar\ and SFR).

		We derive our local comparison samples from the Sloan Digital Sky Survey \citep[SDSS,][]{2000AJ....120.1579Y} Data Release 7 \citep[DR7,][]{2009ApJS..182..543A} catalogue. The emission-line measurements, stellar masses and SFRs are taken from the MPA-JHU catalogues \citep{2003MNRAS.346.1055K,2004MNRAS.351.1151B,2004ApJ...613..898T}. SDSS SFRs are estimated from the \Ha\ luminosities after correcting for aperture loss of the SDSS fibers and dust extinction based on $\Ha/\Hb$. Although the SFRs are based on a Kroupa IMF, SDSS stellar masses are based on a Chabrier IMF. We therefore normalize to a Kroupa IMF (as described for our high-z sample). To estimate the effects of dust extinction and derive ISM conditions we require $\OII \lambda 3727$, $\Hb$, $\OIII \lambda 5007$, $\Ha$, $\NII \lambda 6583$ and $\SII \llambda 6716,6731$ to be detected at a signal-to-noise ratio ($\SN$) $>3$.  We reject AGN, based on the standard optical line ratios, using the \cite{2001ApJ...556..121K} maximum starburst criteria and reduce systematic errors from aperture effects by selecting galaxies at $z>0.04$ \citep{2005PASP..117..227K}. To ensure that we select galaxies representative of the local Universe we limit the redshift to $z\leq 0.1$. These imposed constraints result in a sample of 123652 local star-forming galaxies which we refer to as the ``full local sample''.

		We select three ``matched'' samples from the full local sample by matching an ensemble of local counterparts to each galaxy in our \OII-\Ha\ high-z sample. Our first local comparison sample is matched solely on stellar mass, without applying any constraints to the SFR. We refer to this sample as the \Mstar-matched sample (top row: Fig. \ref{fig:match}). To create the \Mstar-matched sample, we require the stellar mass of the local counterparts to be within $0.2\dex$ of their high-z analogues. Conversely, for our second local comparison sample we require the SFRs of the high-z galaxies and local counterparts to be consistent within $0.2\dex$ but impose no constraints on \Mstar. We refer to the second sample as the SFR-matched sample (middle row: Fig. \ref{fig:match}). We derive our third local sample, the \Mstar-and-SFR-matched sample, by combining constraints on the stellar mass and SFR. To select the \Mstar-and-SFR-matched sample we require that the sSFR of the high-z sample and local counterparts are consistent to within $0.2\dex$ and both the \Mstar\ and SFR are consistent to within $0.3\dex$.

		To ensure that the statistical properties of the matched local and high-z samples are equivalent we select the same number of local counterparts for each high-z galaxy. Although there were more than 50 local galaxies with equivalent \Mstar\ for each high-z galaxy, a greater sample size did not result in a change in the electron density distribution.  We therefore limit the size of our \Mstar-matched sample, by randomly selecting 50 local galaxies for each high-z galaxy. In contrast, the number of local counterparts in both the SFR-matched and the \Mstar-and-SFR-matched sample is limited by the rarity of high SFR galaxies in the local SDSS sample. We only find 7 local counterparts for our highest SFR high-z galaxy and thus select 7 local galaxies at random for the remainder of our sample. Because we impose further constraints to select the \Mstar-and-SFR-matched sample we are limited to 5 local counterparts for each high-z galaxy.

		The three matched local samples have significantly different distributions of \Mstar, SFR and sSFR (see Fig. \ref{fig:match}). Both the \Mstar-matched and SFR-matched local samples have significantly lower sSFRs than our high-z sample, reflecting the evolution of the main star-forming sequence (\Mstar\ vs \SFR) with redshift \citep[e.g.][]{2014ApJS..214...15S}. As shown in Fig. \ref{fig:match} most local galaxies with equivalent stellar masses to galaxies in the high-z sample have lower SFRs. Conversely, most local galaxies with equivalent SFRs to our high-z sample (of which there are far fewer) are more massive than our high-z galaxies. We note that because of the rarity of high SFR local galaxies and the selection criteria imposed, there is significant overlap between the SFR- and the \Mstar-and-SFR- matched samples.

%%%%%%%%%%%%%%%%%%%%%%%%%%%%%%%%%%%%%%%%%%%%%%%%%%%%%%%%%%

\section{Electron Densities}
\label{sec:Eldens}

	The electron density is a useful diagnostic of the pressure and density of gas within star-forming regions. Greater electron densities may help drive the elevated emission-line ratios observed at high redshift by increasing the rate of collisional excitation \citep[e.g.][]{2016ApJ...816...23S,0004-637X-774-2-100,2014ApJ...787..120S}.  The electron density can be estimated using the ratio of an emission-line doublet arising from a single species in which the two energy levels have nearly the same excitation energy but different collisional strengths and radiative probabilities \citep{1987ApJS...63..295V}.  Two electron density sensitive doublets, $\OII \llambda 3726,3729$ and $\SII \llambda 6716,6731$, may be accessed via rest-frame optical spectra.

	% money figure
	\begin{figure*}
	  \begin{center}  % trim={<left> <lower> <right> <upper>}
		    \includegraphics[width=\textwidth,trim={0.5cm 2.0cm 4.5cm 2.5cm},clip]{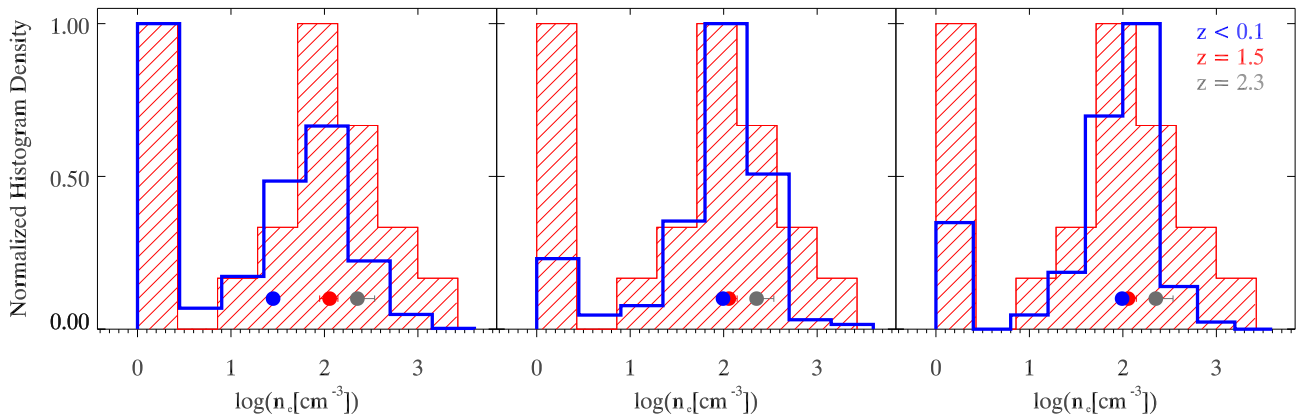}
		  \end{center}
		\caption{Distribution of electron densities for our $z=1.5$ \OII-\Ha\ sample (red line fill), and local comparison samples matched to high-z galaxies for which we determine electron densities (blue outline). The \Mstar-, SFR- and \Mstar-and-SFR- matched local comparison samples are shown in the left, middle and right panels, respectively.  For each panel, the typical electron density of the matched local comparison sample (filled blue circle) is compared to the typical electron density of our \OII-\Ha\ high-z sample (filled red circle) and the electron density of the $z\sim 2.3$ sample from Sanders et. al. (2016) (filled grey circle).
		}
		\label{fig:eldens hist}
	\end{figure*}

	\subsection{Methods} 
		\label{sub:methods}

		We rely on different emission-line doublets to estimate the electron densities of our local and high-z samples because of the difference in resolution elements.  The $z\sim 1.5$ DEIMOS spectra have a spectral resolution of $R\sim 2000$, which is sufficient to fully resolve the \OII \llambda 3726,3729 doublet (separated by $\sim 6.8\ang$ at $z\sim 1.5$). In contrast, local SDSS spectra have a spectral resolution of $R=1800$ at the [OII] wavelength \citep{2000AJ....120.1579Y}, corresponding to a resolution element of $\sim 2.1 \ang$. Because the SDSS spectral resolution cannot fully resolve the 2.78\ang separation of the components of the \OII\ doublet we rely on the  \SII \llambda 6716,6731 doublet to estimate the electron density of the local sample. As shown by \cite{2016ApJ...816...23S}, the electron densities determined from \SII\ and \OII\ for individual HII regions are highly consistent. Because the lines in both doublets are sufficiently close in wavelength, no correction for dust extinction is necessary.

		We calculate the electron densities of our local and high-z samples using the functional form derived in \cite{2016ApJ...816...23S},
		  \begin{align}
	          n_e(R) = \dfrac{cR -ab}{a-R} \, ,
	          \label{eq:e_dens}
	      \end{align}
    	where $R$ is ratio between the peak fluxes of the two emission-line doublet components, $n_e$ is the electron density in \cmcub\ and $a,\,b$ and $c$ are the coefficients which best fit the numerical solutions to the relative populations of the doublets. We provide the coefficients and limiting line ratios derived by \cite{2016ApJ...816...23S} in Table \ref{tab:coefficients}.   

    	The relationships derived by \cite{2016ApJ...816...23S} are the result of a detailed balance of transitions for each of the five energy levels approximated for the O$^{+}$ and S$^{+}$ ions. Calculating the emission-line ratio corresponding to a given density requires accurate transition probabilities and collision strengths. \cite{2016ApJ...816...23S} rely on the most up-to-date atomic data taking transition probabilities for both \OII\ and \SII\ from the NIST MCHF database \citep{nist_fischer_tachiev} and collision strengths from \cite{2007ApJS..171..331T} and \citep{2010ApJS..188...32T} for \OII\ and \SII\ respectively. The methods and atomic data implemented by \cite{2016ApJ...816...23S} are validated by the one-to-one relation between the electron densities of local HII regions derived from \OII\ and \SII.  

    	\begin{table}
      	% \begin{center}
        \centering
        \ra{1.3}
        \caption{Coefficients and limiting line ratios for \OII\ and \SII\ applied to equation \eqref{eq:e_dens}}
        \label{tab:coefficients}
        % \begin{threeparttable}
         \begin{tabular}{c|ccc|cc} %@{}
         \toprule
          Ratio & a & b & c & $R_\mathrm{min}^1$ & $R_\mathrm{max}^2$ \\
         \toprule
              $\OII \lambda 3726/\lambda 3729$ & 0.3771 & 2468 & 638.4 & 0.3839 & 1.4558 \\
              $\SII \lambda 6716/\lambda 6731$ & 0.4315 & 2107 & 627.1 & 0.4375 & 1.4484 \\
         \bottomrule
       \end{tabular}
       \begin{tablenotes}[flushleft]
      %   \small
          \item $^1$ Theoretical minimum line ratio calculated in the high-density limit of $10^5\cmcub$ 
          \item $^2$ Theoretical maximum line ratio calculated in the low-density limit of $1\cmcub$
        \end{tablenotes}
      % \end{threeparttable}
       % \end{center}
      \end{table}

    	The functional form derived by \cite{2016ApJ...816...23S} incorporates a number of assumptions. By using their relation we assume that all of the star-forming galaxies in our samples can be modelled as HII regions consisting of a fully ionized gas with an isobaric density distribution. We thereby assume that the electron density is directly proportional to the HII region pressure and that these regions have an electron temperature of $10^4$K (see e.g., \citealt{0004-637X-647-1-244}, for a discussion). Given the dependence of the collision strength upon temperature this assumption may lead to an over-estimation in the electron density for metal-rich galaxies (and vice versa for metal-poor galaxies). We note that the uncertainty introduced by this assumption is significantly less than the typical measurement error for individual galaxies.   

    % subsection methods (end)

    \subsection{Electron density samples} % (fold)
    	\label{sub:high_z_electron_density_sample}

    	%P1: Careful selection of galaxies 
		Small changes in the line fluxes can have a significant impact upon the inferred electron density, especially at low densities where the line ratio is close to unity. It is therefore crucial that the doublets used to infer electron densities are free of any contamination from sky lines. We visually inspect all of our \OII\ doublets and remove any spectra showing evidence of skyline contamination in the wavelength range of the doublet (starred galaxies in Table \ref{tab:main_sample}).  In addition, we require $\SN>5$ for the line fluxes of both \OII\ components and use the covariance of the line fluxes of the doublets to ensure that we only select galaxies with $\SN>3$ for the flux ratio. 

		Our applied selection cuts result in 57 \OII\ and 21 \OII-\Ha\ high-z galaxies for which we calculate electron densities. For the one galaxy resolved into two separate star-forming regions (Deimos ID: ``D416912'') we separately calculate the electron density of each region and average the result. We find no evidence for AGN contamination in either the \OII\ or \OII-\Ha\ $z\sim 1.5$ samples for which we calculate electron densities, based on the $\OIII/\Hb$ and $\NII/\Ha$ diagnostic line ratios \citep{2001ApJ...556..121K} and lack of X-ray detections. Our \OII-\Ha\ electron density subsample has a median stellar mass of $10^{10.59}\Msun$, a median $\SFR = 28\Msunyr$ and median $\sSFR=0.7\perGyr$ (note that these values differ slightly from the medians for the larger \OII-\Ha\ sample). To ensure a fair comparison, we only present the electron densities for local galaxies matched to the subsample of high-z \OII-\Ha\ detected galaxies for which we estimate electron densities. 
			
    % subsection high_z_electron_density_sample (end)

     \subsection{Electron densities at $z\sim 0$ and $z\sim 1.5$} % (fold)

\begin{figure*}
		      \begin{center}  
		         \includegraphics[width=0.95\textwidth,trim={2.5cm 1cm 0cm 4.5cm},clip]{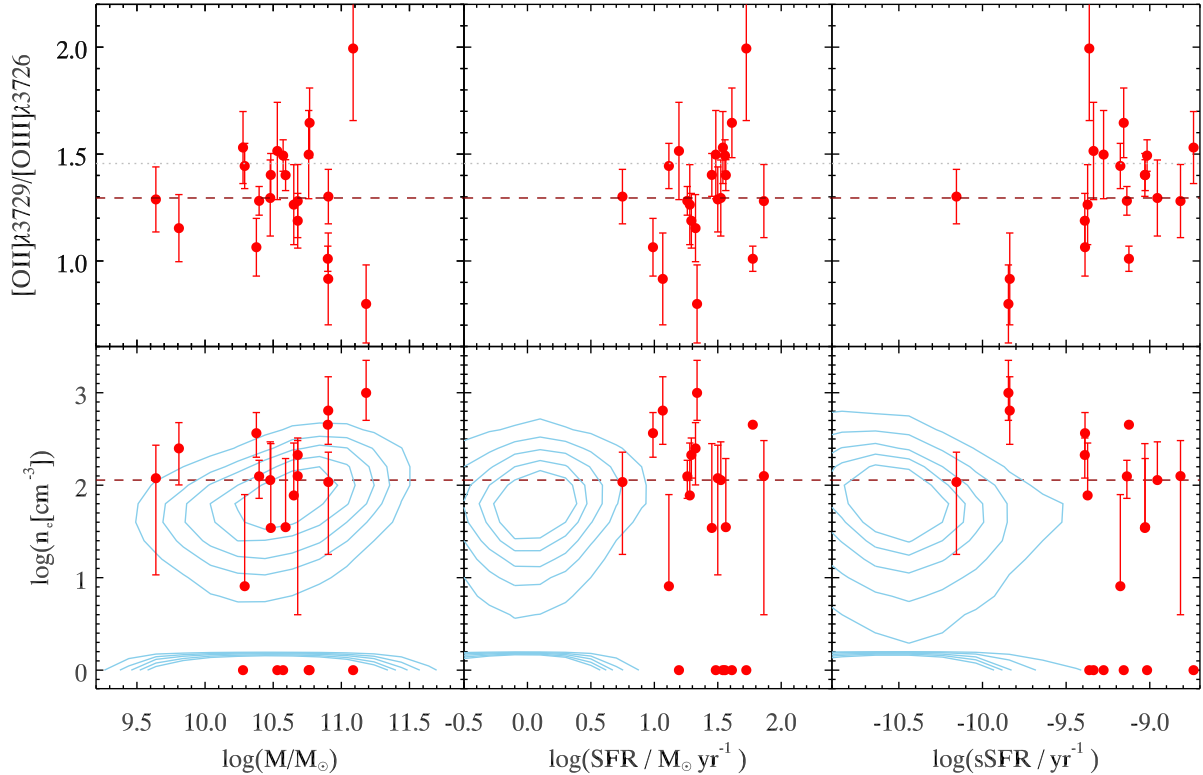}%
		        \end{center} 
		      \caption{$\OII \lambda 3726/\lambda 3729$ (top) and $\log(n_e)$ (bottom) as a function of \Mstar\ (left), \SFR\ (middle) and \sSFR\ (right) for the $z\sim 1.5$ \OII-\Ha\ detected sample. The light blue contours in the bottom row show the regions encompassed by 50\%, 60\%, 70\%, 80\% and 90\% of the full local SDSS sample. The median $\OII \lambda 3726/\lambda 3729$ and typical electron density are  indicated by the red dashed lines in the upper and lower row respectively. The maximum theoretical $\OII \llambda 3726/3729$ is indicated in the top row by the dotted grey lines. }
		      \label{fig:oii and eldens}
		  \end{figure*}

     	We present both the distributions and ``typical'' values of the electron density for each of our samples in Fig. \ref{fig:eldens hist}. Galaxies with line ratios above the theoretical maximum are assigned limits in the low density ($<10\, \cmcub$) regime. To avoid confusion in Fig. \ref{fig:eldens hist} and Fig. \ref{fig:oii and eldens} we assign galaxies with line ratios above the theoretical maximum an electron density of $1\, \cmcub$ (noting that ratios close to the maximum theoretical line ratio can result in densities $<10\, \cmcub$). Although a significant proportion of galaxies fall below the low density limit we find no galaxies with electron densities in the high density regime. 

		Because most samples contain a significant fraction of galaxies below the low density limit, we avoid averaging electron densities. Instead, we determine the ``typical'' electron density of each sample using the median line ratio and applying equation \ref{eq:e_dens}. We estimate the uncertainty on the median line ratio via a resampling technique. For each iteration, we perturb the emission-line ratios according to their uncertainties and take the median of the new sample. We perform this process 1000 times to build a well-sampled distribution of median values. The reported lower and upper uncertainties of the line ratios correspond to the 15.8th and 84.2th percentile values, respectively, of the cumulative distribution function of the median. Lower and upper uncertainties on the typical electron density are determined by converting the uncertainties on the median line ratio to electron densities where the upper (lower) uncertainty in the line ratio corresponds to the lower (upper) uncertainty in the density. The median line ratios and typical electron densities of the local and high-z samples are provided in Table \ref{tab:typical_densities}. 

		\begin{table}
    		\centering
    		\ra{1.3}
			\caption{Median line ratios and typical electron densities of the local and high-z samples}
			\label{tab:typical_densities}
			\begin{tabular}{@{}lcccc@{}}
			\toprule
			\multicolumn{3}{c}{High-z samples}    \\
			\toprule
  			Sample name               &
  			Median      & %\OII \llambda 3726/3729, \SII \llambda 6716/6731
  			$n_e $                   \\
  			&
  			$\OII \lambda 3726/\lambda3729$ &
  			[\cmcub] \\
  			\midrule
  			$\OII$ detected 	& 	$1.32\pm0.02$		& 	$ 90_{-15}^{+17}$	\\
  			$\OII-\Ha$ detected	&	$1.29\pm0.03$ 	&	$114_{-27}^{+28} $ \\
  			\toprule
  			\multicolumn{3}{c}{Local samples}    \\
  			\toprule
  			Sample name               &
  			Median  & %\OII \llambda 3726/3729, \SII \llambda 6716/6731
  			$n_e $                   \\
  			&
  			$\SII \lambda 6716/\lambda 6731$  &
  			[\cmcub] \\
  			\hline
  			Full local sample 	&	$1.4081\pm0.0002$ 	&	$26.8_{-0.2}^{+0.2}$	\\
  			\Mstar-matched 	&	$1.406\pm0.003$	&	$28_{-2}^{+2}$	\\
  			SFR-matched 	&	$1.310\pm0.004$	&	$98_{-4}^{+4}$	\\
  			\Mstar-and-SFR-matched 	&	$1.312\pm0.006$	&	$98_{-5}^{+5}$	\\
  			\bottomrule
		\end{tabular}
		% \end{center}
		\end{table}

		The combined effects of redshift and spectral resolution force us to use different doublets to determine the electron densities of the local and high-z samples. Because [SII] has a significantly lower ionization energy than [OII] (10.36eV and 13.61eV respectively) it can exist at greater nebular distances (Fig. 2, \citealt{1538-3881-139-2-712} and Fig. 4 \citealt{2011MNRAS.417..420M}), sometimes extending well into the diffuse ISM. %beyond the boundary of the HII regions we are attempting to probe. 
		We expect the diffuse ISM to have a negligible effect on the integrated line ratios because we are measuring luminosity-weighted average emission-line spectra, which are dominated by the brightest HII regions.
		Moreover, integrated measurements of HII regions have demonstrated that the \SII\ and \OII\ derived densities are consistent \citep[e.g.][]{2016ApJ...816...23S}, supporting our work. %Confirming the effects of contamination by diffuse emission may be achieved via deeper observations.

		Although individual \SII\ densities could not be derived for our high-z sample, \cite{2016arXiv160406802K} have determined the \SII\ ratio of the stacked spectra of 701 FMOS galaxies with \Ha\ detections at $\SN>3$, of which our \OII-\Ha\ sample is a subsample. Their stacked spectra yield an average \SII\ ratio of $1.21\pm 0.1$ which translates to an electron density of $193_{-93}^{+121}\cmcub$, using the methods presented here. 
		This electron density estimate is higher than the typical electron density of our \OII-\Ha\ sample ($114_{-27}^{+28} \cmcub$), but consistent within the uncertainties. If this offset applies to the individual galaxies of our \OII-\Ha\ sample, it would enhance the difference between the high-z and \Mstar-matched local sample. However, it is still unclear how well the electron density of the stacked dataset reflects the mean electron density of the FMOS galaxies, especially for our sub-sample.

    % subsection typical_electron_density_at_low_and_high_z (end) 

	\subsection{Electron Density vs. Redshift} % (fold)
		\label{sub:electron_denisty_vs_redshift}

		Our results suggest that the elevated electron densities measured at high redshift are a consequence of probing populations of galaxies with far greater star formation activity than is typical of the local Universe. We measure a typical electron density of $27 \cmcub$ for our full sample of 123652 local star-forming galaxies, consistent with the $\sim 20\cmcub$ found by \cite{2016ApJ...816...23S} and \cite{2016ApJ...822...62B}. As found in previous high-redshift studies \citep{2016ApJ...816...23S,2014ApJ...787..120S,2014ApJ...795..165S,2014ApJ...785..153M}, the typical electron density of our high-z sample is significantly greater ($ \sim 5 \times$) than that of the local galaxy population. The same offset is recovered for local and high-z samples matched in stellar mass only. However, we find no significant difference between the typical electron densities of local and high-z samples with equivalent SFRs ($98_{-5}^{+5}\cmcub$ vs. $114_{-27}^{+28}\cmcub$ respectively). 

		Our findings appear to contradict the work of \cite{2014ApJ...787..120S}, who recover a significant difference between the electron densities of their \Mstar-and-SFR-matched local and high-z samples. The conflicting results are most likely to be the result of the different methods employed to estimate electron densities. \cite{2014ApJ...787..120S} measure the electron densities of their local sample using the \SII\ ratio and derive the electron density for each corresponding high-z galaxy by applying a scaling factor based on the ratio between the ionization parameters of the matched local and high-z galaxies. In contrast, we directly determine the electron density for the local and high-z samples, based on measured doublet ratios. We separately investigate the ionization parameter in Kaasinen et al (in prep.) to determine whether we find the same dependence on star formation rate as for the electron density.

	% subsection electron_density_vs_redshift (end)

	\subsection{Electron density vs global galaxy properties} 
	\label{sub:electron_density_vs_global_galaxy_properties}

		Our work indicates that the previously observed evolution in electron density is related to the evolution of SFR rather than \Mstar. Recent high-redshift studies which measure high electron densities, probe samples of galaxies with higher SFRs than typically found at $z<1.5$.  Both \cite{2014ApJ...795..165S} and \cite{2014ApJ...785..153M} estimate a typical electron density of $\sim 243 \cmcub$ for their $z\sim 2.3$ and $z\sim 1.85$ samples which have median SFRs of $20$ and $25\Msunyr$ respectively. Similarly, \cite{2016ApJ...816...23S} measure a typical electron density of $\sim 225 \cmcub$ for their $z\sim 2.3$ \OII\ detected sample which has a median SFR of $\sim 30 \Msunyr$. \cite{2015MNRAS.451.1284S} measure a typical electron density of $\sim 290 \cmcub$ for their sample of 14 \Ha\ emitters at $z\sim 2.5$ for which the median SFR is $\sim 100 \Msunyr$. Although the SFRs of these high-z samples differ significantly from those of the local comparison samples used, the stellar mass ranges are directly comparable. Thus, it would appear as though variations in electron density are driven mainly by differences in SFR. 

		We investigate how the doublet ratios and electron density vary with global galaxy properties both graphically (Fig. \ref{fig:oii and eldens}) and by performing a Spearman Rank correlation test for each parameter space.  To determine the extent to which the variables are related we measure the strength of the monotonic relationship for each pairing via the Spearman rank coefficient, $\rho_s$, and consider the significance of the correlation via $\alpha$, the likelihood of $\rho_s$ being found by chance if the two variables are uncorrelated. For our high-z \OII-\Ha\ sample there is neither graphical nor statistical evidence for any correlations (i.e. $\alpha>0.15$ in all cases). These findings contradict the significant correlation ($4\sigma $ level) between electron density and sSFR found by \cite{2015MNRAS.451.1284S}. However, given the limited range in global properties and extent of the uncertainty for each individual measurement, our high-z sample is likely to be too small to recover any underlying correlations. 
 
\begin{figure*}
		  \begin{center}  
			    \includegraphics[width=\textwidth,trim={1.5cm 0.3cm 2.5cm 0.5cm},clip]{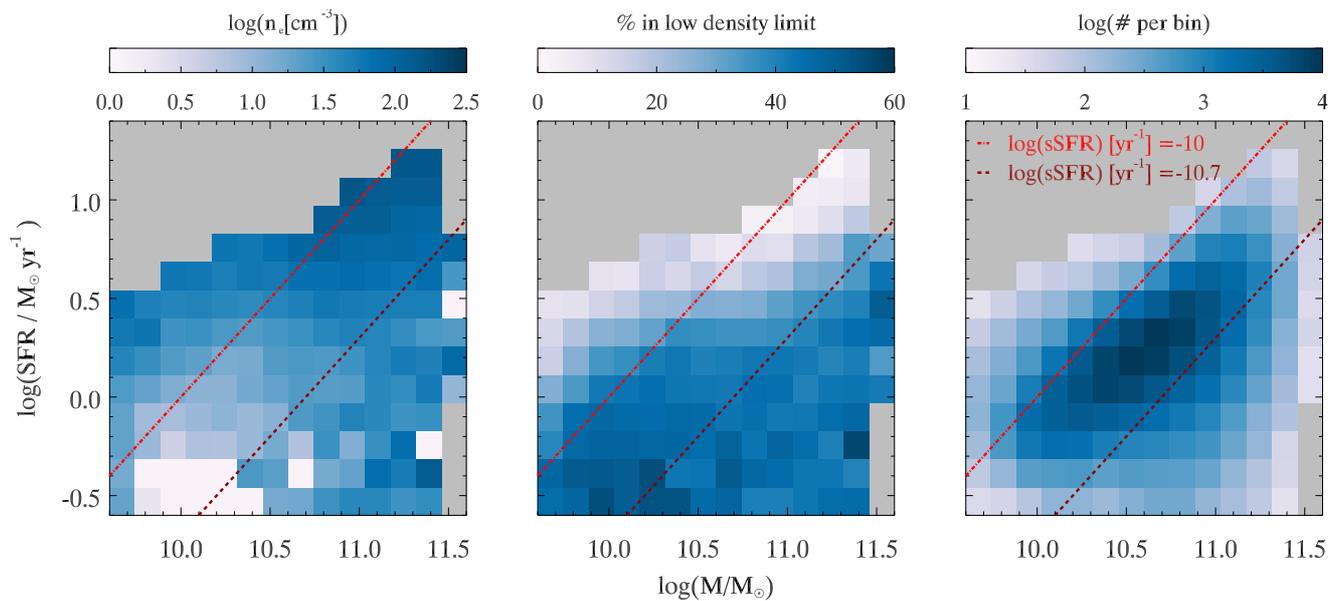}
			  \end{center}
			\caption{Electron density as a function of SFR and \Mstar\ in local SDSS galaxies. Locii of constant sSFR are overplotted in orange and dark red for $\log (\sSFR / \peryr) > -10$ and $\log (\sSFR / \peryr) > -11$ respectively. Left panel: typical electron density for each bin of $\geq 20$ galaxies. Middle panel: fraction of galaxies in each bin with $\SII \lambda 6716/\lambda 6731$ above the theoretical maximum i.e. fraction of galaxies in the low density regime. Right panel: Number of galaxies per bin.}
			\label{fig:eldens glob}
		\end{figure*}

		Unlike the high-z sample, the local samples exhibit weak but significant correlations. For each of the local samples (listed in Table \ref{tab:typical_densities}) we find a weak negative Spearman's correlation ($\rho_s\sim -0.1$) between the \SII\ line ratio and \Mstar, which translates into a weak positive correlation between electron density and stellar mass ($\rho_s\sim 0.2$). Additionally, each of the local samples exhibits a weak, negative correlation between the \SII\ ratio and SFR which translates to a weak positive correlation between the electron density and SFR. The $\rho_s$ values describing the relationship between electron density and SFR are greater for the local samples matched in SFR ($\rho_s\sim 0.4$) than for the full and \Mstar-matched samples ($\rho_s\sim 0.15$). Further differences between samples are apparent when investigating the relationship between electron density and sSFR. Both the full and \Mstar-matched local samples exhibit no correlation between the electron density and sSFR whereas the samples matched in SFR exhibit weak positive correlations ($0.1<\rho_s<0.3$). The weakness of the monotonic relationships and differences between samples are reflective of the contours for the full local sample shown in Fig. \ref{fig:eldens glob}. At high SFR ($\log(\SFR/\Msunyr)>0.5$) and sSFR ($\log(\sSFR/\peryr)>-10$) the range of likely electron densities is smaller and offset to higher values.

		We extend our investigation into the dependence of electron density on global properties using the full sample of 123652 local star-forming galaxies. Rather than separately investigating the correlation between electron density and each global property, we compare the electron densities of different bins of \SFR\ and \Mstar. For each bin we compute the typical electron density (as described in Sec. \ref{sub:methods}), determine the fraction of galaxies in the low density regime and count the total number of galaxies to account for sample characteristics (Fig. \ref{fig:eldens glob}: left, middle and right panels respectively). We also show lines of constant \sSFR\ (Fig. \ref{fig:eldens glob}: red dot dashed and dashed lines). We require at least 20 galaxies per SFR and \Mstar\ bin but impose no upper limit on the number. Bins where $-10 < \log (\sSFR / \peryr) < -10.7$ and $10.2< \log(\Mstar/\Msun) <11.2$ contain the greatest number of galaxies, reflecting the characteristics of SDSS as well as our selection criteria. 

		The electron density of the bins of local galaxies appears to be more strongly dependent on the star-formation rate than the stellar mass. For each \Mstar\ bin the typical electron density increases with SFR above $\sim \log (\sSFR / \peryr) > -10.7$ (left panel, Fig. \ref{fig:eldens glob}). Furthermore, the fraction of galaxies with electron densities in the low density regime decreases dramatically with increasing SFR, above $\log (\sSFR / \peryr) > -10.7$. This dependence on SFR (and sSFR) is also apparent for the matched local samples. Approximately $40\%$ of both the full local sample and \Mstar-matched sample have electron densities in the low density limit, whereas $\sim 15\%$ of the samples matched in SFR exhibit electron densities in the low density regime. Thus, the previously observed increase in electron density at high redshift may be driven by the decreasing fraction of galaxies within the low density regime for populations with higher SFRs. 

		The dependence of the electron density of a sample on the SFR may reflect a correlation between electron density and star-formation rate density. Because it is observationally challenging to determine the SFR volume density ($\rho_\SFR$) for galaxies other than our own, the SFR relative to the size of a galaxy is commonly measured via the SFR surface density ($\Sigma_\SFR$). Recently, studies have found strong evidence for a correlation between the global $\Sigma_\SFR$ and electron density of star-forming galaxies \citep{2016ApJ...822...62B,2015MNRAS.451.1284S}. Although we lack the spatial information required to measure the size, and therefore $\Sigma_\SFR$, for our $z\sim 1.5$ sample we note that our high-z sample is unlikely to be biased towards greater galaxy sizes than the local population \citep[e.g.][]{2014ApJ...788...28V}. We therefore expect our high-z sample to have a greater average $\Sigma_\SFR$ than the local population and at least equivalent $\Sigma_\SFR$ to the local samples matched in SFR. The same argument should extend to $\rho_\SFR$ since the growth in galaxy size should be independent of the viewing angle.

		There are a number of physical mechanisms by which a high star formation rate density may result in an enhanced average electron density. A higher SFR density means an increased number of massive, young stars per unit volume. The increased number density of massive, young, stars results in a greater energy input to HII regions, via processes such as stellar winds and shocks, increasing both their pressure and electron density \citep{2009ApJ...703.1352K,2008ApJS..176..438G}. Additionally, the observed correlation between the star formation rate density and electron density may ensue as a result of the direct scaling between the gas surface density ($\Sigma_\mathrm{gas}$) and $\Sigma_\SFR$ \citep[e.g.][]{2012ARA&amp;A..50..531K,1998ApJ...498..541K}. Increased molecular gas densities may lead to higher atomic hydrogen densities within HII regions, which are embedded in molecular gas clouds. Because the electron density scales with the number density of atomic hydrogen \citep[e.g.][]{2003adu..book.....D,0004-637X-774-2-100,0004-637X-647-1-244,0067-0049-167-2-177} it may also be sensitive to the gas density. Thus, the high $\Sigma_\mathrm{gas}$ of high-z galaxies \citep[e.g.][]{2010Natur.463..781T,2013ApJ...773...68G} would be expected to increase both the SFR and electron density. 
 
		In the previous arguments we assumed that the observed emission-line fluxes stem predominantly from HII regions and that the estimated electron densities are therefore reflective of the average HII region electron density for each galaxy. But, studies on local galaxies show that 20-40\% of the total galactic \Ha\ luminosity stems from the diffuse ISM, and possibly a larger fraction of the [OII] and [SII] emission  \citep[][and references therein]{2003adu..book.....D,2009RvMP...81..969H}. %Additionally, the diffuse ISM still contains a  significant proportion of ionized oxygen and sulfur. We are therefore likely to be probing the average electron density of both the HII regions and diffuse ISM. 
		Most studies of the diffuse ISM \citep[see][]{2009RvMP...81..969H} do not report any evidence for a correlation between SFR (or \Ha\ luminosity) and the fraction of the emission stemming from the diffuse ISM,  suggesting that the diffuse emission may affect our high-z and local samples. %It therefore remains unclear to what extent the proportion of diffuse gas within a galaxy affects emission line properties, especially at high redshift. 
		Confirming the effects of contamination by diffuse emission will only be achieved via deeper observations and high physical resolution IFU observations of both high- and low-z galaxies.

%%%%%%%%%%%%%%%%%%%%%%%%%%%%%%%%%%

\section{Summary}
\label{sec:Summary}

	We have presented the data and first results from the COSMOS \OII\ survey.  Our survey is designed to complement the FMOS-COSMOS survey \citep{2013ApJ...777L...8K,2014arXiv1409.0447S,0004-637X-792-1-75,2016arXiv160406802K} by measuring the flux of the \OII\ doublet for galaxies with H- and J-long observations. As shown in this work, our data represent a critical resource for probing the ionization state of star-forming regions at $z\sim 1.5$. Without the \OII\ doublet we cannot accurately diagnose the electron density, ionization parameter or metallicity of star-forming regions.
	
	We have investigated the average electron density of star-forming galaxies at $z\sim 1.5$ and made comparisons to local star-forming galaxies. Of the 103 galaxies at $z\sim 1.5$ with significant \OII\ detections, a subsample of 46 have measured \Ha. We use this subsample to explore the effects of \Mstar\ and \SFR\ on electron density. To measure the electron density we select a subsample with high S/N for both components of the \OII\ doublet and for which there is no evidence of skyline contamination. We measure a median $\OII \lambda 3726/\lambda 3729$ of $1.29\pm0.03$ for our $z\sim 1.5$ \OII-\Ha\ detected sample, corresponding to a typical electron density of $114_{-27}^{+28}\cmcub$. This typical electron density is consistent with recent high-z ($1.8<z<2.5$) studies as well as the electron density found by \cite{2016arXiv160406802K} for the parent \Ha\ detected FMOS-COSMOS sample.

	We find strong evidence that the high electron densities measured for high-redshift galaxies are the result of the evolving global star-formation rate. Our $z\sim 1.5$ sample exhibits a typical electron density $\sim 5$ times that of the ``typical'' local star-forming galaxy population and a local sample matched in stellar mass. But, we find no such evolution in electron density when comparing local and high-z galaxies with the same SFR. The dependence of electron density on SFR is also evident within the local sample, for which we plot the typical electron density and fraction of galaxies in the low density regime as a function of \Mstar\ and SFR. The dependence we have uncovered may be the result of a correlation with the star-formation rate surface density, reported by other studies \citep[e.g.][]{2015MNRAS.451.1284S}. However, further studies of spatially resolved galaxies at $0<z<2$ are required to confirm this.

	Our findings indicate that the enhanced emission-line ratios observed at high redshift are, at least partly, the result of probing populations of galaxies with higher \SFR\ and \sSFR\ than are typical of the local Universe. Higher electron densities serve to increase emission-line ratios such as \OIII/\Hb, \citep[e.g.][]{0004-637X-774-2-100}. Thus, the increased star-formation activity at high redshift may be reflected in enhanced ratios. % High redshift studies which find elevated emission line ratios and electron densities \citep[e.g.][]{2016ApJ...816...23S,2015MNRAS.451.1284S,2014ApJ...787..120S,2014ApJ...795..165S}, probe populations of galaxies with significantly higher \SFR\ and \sSFR\ than observed for the local galaxy population. 
	The proposed correlation between emission-line ratios and \sSFR\ has only been investigated recently. \cite{2016ApJ...822...62B} find evidence that local galaxies with \OIII/\Hb\ ratios equivalent to galaxies at $z\sim 2$ have significantly higher sSFRs and electron densities than the rest of the local star-forming population. Similarly, \cite{2016arXiv160601259D} show that elevated \OIII/\Hb\ ratios are partly driven by high \sSFR, regardless of the cosmic epoch. 

	The exact connection between global galaxy properties and the conditions within star-forming regions remains unclear. Although the electron density is one of the key physical parameters of star-forming regions, it does not convey the full picture. To fully probe the physical conditions of star-forming regions at $z\sim 1.5$ we must also investigate their metallicities and ionization parameters. Previous studies, which sample galaxies with high star-formation activity, find evidence for increased ionization parameters \citep{2015MNRAS.451.1284S,2014ApJ...795..165S,2016ApJ...816...23S}. We will investigate the observed evolution of the ionization parameter and its dependence on global galaxy properties in Kaasinen et al. (in prep) using the high-z sample presented in this work.

\section*{Acknowledgements}

	We acknowledge the data, feedback and support provided by the FMOS-COMOS team, in particular Kashino Daichi, Jabran Zahid and John Silverman. B.G. gratefully acknowledges the support of the Australian Research Council as the recipient of a Future Fellowship (FT140101202). LK gratefully acknowledges support from an ARC Laureate Fellowship (FL150100113). We also thank the anonymous referee for the insightful comments which greatly improved this paper.

	The data presented herein were obtained at the W.M. Keck Observatory, which is operated as a scientific partnership among the California Institute of Technology, the University of California and the National Aeronautics and Space Administration. The Observatory was made possible by the generous financial support of the W.M. Keck Foundation. We wish to recognize and acknowledge the very significant cultural role and reverence that the summit of Mauna Kea has always had within the indigenous Hawaiian community.  We are fortunate to have the opportunity to conduct observations from this mountain. 

%%%%%%%%%%%%%%%%%%%%%%%%%%%%%%%%%%%%%%%%%%%%%%%%%%



\bibliographystyle{mnras}
\bibliography{Kaasinen_bib}

\begin{thebibliography}{}
\makeatletter
\relax
\def\mn@urlcharsother{\let\do\@makeother \do\$\do\&\do\#\do\^\do\_\do\%\do\~}
\def\mn@doi{\begingroup\mn@urlcharsother \@ifnextchar [ {\mn@doi@}
  {\mn@doi@[]}}
\def\mn@doi@[#1]#2{\def\@tempa{#1}\ifx\@tempa\@empty \href
  {http://dx.doi.org/#2} {doi:#2}\else \href {http://dx.doi.org/#2} {#1}\fi
  \endgroup}
\def\mn@eprint#1#2{\mn@eprint@#1:#2::\@nil}
\def\mn@eprint@arXiv#1{\href {http://arxiv.org/abs/#1} {{\tt arXiv:#1}}}
\def\mn@eprint@dblp#1{\href {http://dblp.uni-trier.de/rec/bibtex/#1.xml}
  {dblp:#1}}
\def\mn@eprint@#1:#2:#3:#4\@nil{\def\@tempa {#1}\def\@tempb {#2}\def\@tempc
  {#3}\ifx \@tempc \@empty \let \@tempc \@tempb \let \@tempb \@tempa \fi \ifx
  \@tempb \@empty \def\@tempb {arXiv}\fi \@ifundefined
  {mn@eprint@\@tempb}{\@tempb:\@tempc}{\expandafter \expandafter \csname
  mn@eprint@\@tempb\endcsname \expandafter{\@tempc}}}

\bibitem[\protect\citeauthoryear{{Abazajian} et~al.,}{{Abazajian}
  et~al.}{2009}]{2009ApJS..182..543A}
{Abazajian} K.~N.,  et~al., 2009, \mn@doi [\apjs]
  {10.1088/0067-0049/182/2/543}, \href
  {http://adsabs.harvard.edu/abs/2009ApJS..182..543A} {182, 543}

\bibitem[\protect\citeauthoryear{{Adelberger}, {Steidel}, {Shapley}, {Hunt},
  {Erb}, {Reddy}  \& {Pettini}}{{Adelberger}
  et~al.}{2004}]{2004ApJ...607..226A}
{Adelberger} K.~L.,  {Steidel} C.~C.,  {Shapley} A.~E.,  {Hunt} M.~P.,  {Erb}
  D.~K.,  {Reddy} N.~A.,   {Pettini} M.,  2004, \mn@doi [\apj]
  {10.1086/383221}, \href {http://adsabs.harvard.edu/abs/2004ApJ...607..226A}
  {607, 226}

\bibitem[\protect\citeauthoryear{{Arnouts} \& {Ilbert}}{{Arnouts} \&
  {Ilbert}}{2011}]{2011ascl.soft08009A}
{Arnouts} S.,  {Ilbert} O.,  2011, {LePHARE: Photometric Analysis for Redshift
  Estimate}, Astrophysics Source Code Library (\mn@eprint {ascl} {1108.009})

\bibitem[\protect\citeauthoryear{{Bian} et~al.,}{{Bian}
  et~al.}{2010}]{2010ApJ...725.1877B}
{Bian} F.,  et~al., 2010, \mn@doi [\apj] {10.1088/0004-637X/725/2/1877}, \href
  {http://adsabs.harvard.edu/abs/2010ApJ...725.1877B} {725, 1877}

\bibitem[\protect\citeauthoryear{{Bian}, {Kewley}, {Dopita}  \&
  {Juneau}}{{Bian} et~al.}{2016}]{2016ApJ...822...62B}
{Bian} F.,  {Kewley} L.~J.,  {Dopita} M.~A.,   {Juneau} S.,  2016, \mn@doi
  [\apj] {10.3847/0004-637X/822/2/62}, \href
  {http://adsabs.harvard.edu/abs/2016ApJ...822...62B} {822, 62}

\bibitem[\protect\citeauthoryear{{Brinchmann}, {Charlot}, {White}, {Tremonti},
  {Kauffmann}, {Heckman}  \& {Brinkmann}}{{Brinchmann}
  et~al.}{2004}]{2004MNRAS.351.1151B}
{Brinchmann} J.,  {Charlot} S.,  {White} S.~D.~M.,  {Tremonti} C.,  {Kauffmann}
  G.,  {Heckman} T.,   {Brinkmann} J.,  2004, \mn@doi [\mnras]
  {10.1111/j.1365-2966.2004.07881.x}, \href
  {http://adsabs.harvard.edu/abs/2004MNRAS.351.1151B} {351, 1151}

\bibitem[\protect\citeauthoryear{{Brinchmann}, {Pettini}  \&
  {Charlot}}{{Brinchmann} et~al.}{2008}]{2008MNRAS.385..769B}
{Brinchmann} J.,  {Pettini} M.,   {Charlot} S.,  2008, \mn@doi [\mnras]
  {10.1111/j.1365-2966.2008.12914.x}, \href
  {http://cdsads.u-strasbg.fr/abs/2008MNRAS.385..769B} {385, 769}

\bibitem[\protect\citeauthoryear{{Cardelli}, {Clayton}  \& {Mathis}}{{Cardelli}
  et~al.}{1989}]{1989ApJ...345..245C}
{Cardelli} J.~A.,  {Clayton} G.~C.,   {Mathis} J.~S.,  1989, \mn@doi [\apj]
  {10.1086/167900}, \href {http://adsabs.harvard.edu/abs/1989ApJ...345..245C}
  {345, 245}

\bibitem[\protect\citeauthoryear{{Cooper}, {Newman}, {Davis}, {Finkbeiner}  \&
  {Gerke}}{{Cooper} et~al.}{2012}]{2012ascl.soft03003C}
{Cooper} M.~C.,  {Newman} J.~A.,  {Davis} M.,  {Finkbeiner} D.~P.,   {Gerke}
  B.~F.,  2012, {spec2d: DEEP2 DEIMOS Spectral Pipeline}, Astrophysics Source
  Code Library (\mn@eprint {ascl} {1203.003})

\bibitem[\protect\citeauthoryear{{Cowie}, {Barger}  \& {Songaila}}{{Cowie}
  et~al.}{2016}]{2016ApJ...817...57C}
{Cowie} L.~L.,  {Barger} A.~J.,   {Songaila} A.,  2016, \mn@doi [\apj]
  {10.3847/0004-637X/817/1/57}, \href
  {http://adsabs.harvard.edu/abs/2016ApJ...817...57C} {817, 57}

\bibitem[\protect\citeauthoryear{{Cullen}, {Cirasuolo}, {McLure}, {Dunlop}  \&
  {Bowler}}{{Cullen} et~al.}{2014}]{2014MNRAS.440.2300C}
{Cullen} F.,  {Cirasuolo} M.,  {McLure} R.~J.,  {Dunlop} J.~S.,   {Bowler}
  R.~A.~A.,  2014, \mn@doi [\mnras] {10.1093/mnras/stu443}, \href
  {http://adsabs.harvard.edu/abs/2014MNRAS.440.2300C} {440, 2300}

\bibitem[\protect\citeauthoryear{{Daddi} et~al.,}{{Daddi}
  et~al.}{2007}]{2007ApJ...670..156D}
{Daddi} E.,  et~al., 2007, \mn@doi [\apj] {10.1086/521818}, \href
  {http://adsabs.harvard.edu/abs/2007ApJ...670..156D} {670, 156}

\bibitem[\protect\citeauthoryear{{Dickey} et~al.,}{{Dickey}
  et~al.}{2016}]{2016arXiv160601259D}
{Dickey} C.~M.,  et~al., 2016, preprint, \href
  {http://adsabs.harvard.edu/abs/2016arXiv160601259D} {} (\mn@eprint {arXiv}
  {1606.01259})

\bibitem[\protect\citeauthoryear{{Dickinson}, {Papovich}, {Ferguson}  \&
  {Budav{\'a}ri}}{{Dickinson} et~al.}{2003}]{2003ApJ...587...25D}
{Dickinson} M.,  {Papovich} C.,  {Ferguson} H.~C.,   {Budav{\'a}ri} T.,  2003,
  \mn@doi [\apj] {10.1086/368111}, \href
  {http://adsabs.harvard.edu/abs/2003ApJ...587...25D} {587, 25}

\bibitem[\protect\citeauthoryear{{Dopita} \& {Sutherland}}{{Dopita} \&
  {Sutherland}}{2003}]{2003adu..book.....D}
{Dopita} M.~A.,  {Sutherland} R.~S.,  2003, {Astrophysics of the diffuse
  universe}

\bibitem[\protect\citeauthoryear{Dopita et~al.,}{Dopita
  et~al.}{2006a}]{0067-0049-167-2-177}
Dopita M.~A.,  et~al., 2006a, \apjsupp, 167, 177

\bibitem[\protect\citeauthoryear{Dopita et~al.,}{Dopita
  et~al.}{2006b}]{0004-637X-647-1-244}
Dopita M.~A.,  et~al., 2006b, \apj, 647, 244

\bibitem[\protect\citeauthoryear{{Drory}, {Salvato}, {Gabasch}, {Bender},
  {Hopp}, {Feulner}  \& {Pannella}}{{Drory} et~al.}{2005}]{2005ApJ...619L.131D}
{Drory} N.,  {Salvato} M.,  {Gabasch} A.,  {Bender} R.,  {Hopp} U.,  {Feulner}
  G.,   {Pannella} M.,  2005, \mn@doi [\apjl] {10.1086/428044}, \href
  {http://adsabs.harvard.edu/abs/2005ApJ...619L.131D} {619, L131}

\bibitem[\protect\citeauthoryear{{Elbaz} et~al.,}{{Elbaz}
  et~al.}{2011}]{2011A&amp;A...533A.119E}
{Elbaz} D.,  et~al., 2011, \mn@doi [\aap] {10.1051/0004-6361/201117239}, \href
  {http://adsabs.harvard.edu/abs/2011A%26A...533A.119E} {533, A119}

\bibitem[\protect\citeauthoryear{{Faber} et~al.,}{{Faber}
  et~al.}{2003}]{2003SPIE.4841.1657F}
{Faber} S.~M.,  et~al., 2003, in {Iye} M.,  {Moorwood} A.~F.~M.,  eds,
  \procspie Vol. 4841, Instrument Design and Performance for Optical/Infrared
  Ground-based Telescopes. pp 1657--1669, \mn@doi{10.1117/12.460346}

\bibitem[\protect\citeauthoryear{{Fischer}}{{Fischer}}{2014}]{nist_fischer_tachiev}
{Fischer} C.~F. \&~{Tachiev} G.,  2014, \url {http://physics.nist.gov/mchf}

\bibitem[\protect\citeauthoryear{{Fontana} et~al.,}{{Fontana}
  et~al.}{2003}]{2003ApJ...594L...9F}
{Fontana} A.,  et~al., 2003, \mn@doi [\apjl] {10.1086/378489}, \href
  {http://adsabs.harvard.edu/abs/2003ApJ...594L...9F} {594, L9}

\bibitem[\protect\citeauthoryear{{Genzel} et~al.,}{{Genzel}
  et~al.}{2013}]{2013ApJ...773...68G}
{Genzel} R.,  et~al., 2013, \mn@doi [\apj] {10.1088/0004-637X/773/1/68}, \href
  {http://adsabs.harvard.edu/abs/2013ApJ...773...68G} {773, 68}

\bibitem[\protect\citeauthoryear{{Groves}, {Heckman}  \& {Kauffmann}}{{Groves}
  et~al.}{2006}]{2006MNRAS.371.1559G}
{Groves} B.~A.,  {Heckman} T.~M.,   {Kauffmann} G.,  2006, \mn@doi [\mnras]
  {10.1111/j.1365-2966.2006.10812.x}, \href
  {http://adsabs.harvard.edu/abs/2006MNRAS.371.1559G} {371, 1559}

\bibitem[\protect\citeauthoryear{{Groves}, {Dopita}, {Sutherland}, {Kewley},
  {Fischera}, {Leitherer}, {Brandl}  \& {van Breugel}}{{Groves}
  et~al.}{2008}]{2008ApJS..176..438G}
{Groves} B.,  {Dopita} M.~A.,  {Sutherland} R.~S.,  {Kewley} L.~J.,  {Fischera}
  J.,  {Leitherer} C.,  {Brandl} B.,   {van Breugel} W.,  2008, \mn@doi [\apjs]
  {10.1086/528711}, \href {http://adsabs.harvard.edu/abs/2008ApJS..176..438G}
  {176, 438}

\bibitem[\protect\citeauthoryear{{Haffner} et~al.,}{{Haffner}
  et~al.}{2009}]{2009RvMP...81..969H}
{Haffner} L.~M.,  et~al., 2009, \mn@doi [Reviews of Modern Physics]
  {10.1103/RevModPhys.81.969}, \href
  {http://adsabs.harvard.edu/abs/2009RvMP...81..969H} {81, 969}

\bibitem[\protect\citeauthoryear{{Hainline}, {Shapley}, {Kornei}, {Pettini},
  {Buckley-Geer}, {Allam}  \& {Tucker}}{{Hainline}
  et~al.}{2009a}]{2009ApJ...701...52H}
{Hainline} K.~N.,  {Shapley} A.~E.,  {Kornei} K.~A.,  {Pettini} M.,
  {Buckley-Geer} E.,  {Allam} S.~S.,   {Tucker} D.~L.,  2009a, \mn@doi [\apj]
  {10.1088/0004-637X/701/1/52}, \href
  {http://adsabs.harvard.edu/abs/2009ApJ...701...52H} {701, 52}

\bibitem[\protect\citeauthoryear{Hainline, Shapley, Kornei, Pettini,
  Buckley-Geer, Allam  \& Tucker}{Hainline et~al.}{2009b}]{0004-637X-701-1-52}
Hainline K.~N.,  Shapley A.~E.,  Kornei K.~A.,  Pettini M.,  Buckley-Geer E.,
  Allam S.~S.,   Tucker D.~L.,  2009b, \apj, 701, 52

\bibitem[\protect\citeauthoryear{{Hao}, {Kennicutt}, {Johnson}, {Calzetti},
  {Dale}  \& {Moustakas}}{{Hao} et~al.}{2011}]{2011ApJ...741..124H}
{Hao} C.-N.,  {Kennicutt} R.~C.,  {Johnson} B.~D.,  {Calzetti} D.,  {Dale}
  D.~A.,   {Moustakas} J.,  2011, \mn@doi [\apj] {10.1088/0004-637X/741/2/124},
  \href {http://adsabs.harvard.edu/abs/2011ApJ...741..124H} {741, 124}

\bibitem[\protect\citeauthoryear{{Hayashi} et~al.,}{{Hayashi}
  et~al.}{2015}]{2015PASJ...67...80H}
{Hayashi} M.,  et~al., 2015, \mn@doi [\pasj] {10.1093/pasj/psv041}, \href
  {http://adsabs.harvard.edu/abs/2015PASJ...67...80H} {67, 80}

\bibitem[\protect\citeauthoryear{{Holden} et~al.,}{{Holden}
  et~al.}{2016}]{2016ApJ...820...73H}
{Holden} B.~P.,  et~al., 2016, \mn@doi [\apj] {10.3847/0004-637X/820/1/73},
  \href {http://adsabs.harvard.edu/abs/2016ApJ...820...73H} {820, 73}

\bibitem[\protect\citeauthoryear{{Hopkins} \& {Beacom}}{{Hopkins} \&
  {Beacom}}{2006}]{2006ApJ...651..142H}
{Hopkins} A.~M.,  {Beacom} J.~F.,  2006, \mn@doi [\apj] {10.1086/506610}, \href
  {http://adsabs.harvard.edu/abs/2006ApJ...651..142H} {651, 142}

\bibitem[\protect\citeauthoryear{{Ilbert} et~al.,}{{Ilbert}
  et~al.}{2013}]{2013A&amp;A...556A..55I}
{Ilbert} O.,  et~al., 2013, \mn@doi [\aap] {10.1051/0004-6361/201321100}, \href
  {http://adsabs.harvard.edu/abs/2013A%26A...556A..55I} {556, A55}

\bibitem[\protect\citeauthoryear{{Ilbert} et~al.,}{{Ilbert}
  et~al.}{2015}]{2015A&amp;A...579A...2I}
{Ilbert} O.,  et~al., 2015, \mn@doi [\aap] {10.1051/0004-6361/201425176}, \href
  {http://adsabs.harvard.edu/abs/2015A%26A...579A...2I} {579, A2}

\bibitem[\protect\citeauthoryear{{Juneau} et~al.,}{{Juneau}
  et~al.}{2014}]{2014ApJ...788...88J}
{Juneau} S.,  et~al., 2014, \mn@doi [\apj] {10.1088/0004-637X/788/1/88}, \href
  {http://adsabs.harvard.edu/abs/2014ApJ...788...88J} {788, 88}

\bibitem[\protect\citeauthoryear{{Kashino} et~al.,}{{Kashino}
  et~al.}{2013}]{2013ApJ...777L...8K}
{Kashino} D.,  et~al., 2013, \mn@doi [\apjl] {10.1088/2041-8205/777/1/L8},
  \href {http://adsabs.harvard.edu/abs/2013ApJ...777L...8K} {777, L8}

\bibitem[\protect\citeauthoryear{{Kashino} et~al.,}{{Kashino}
  et~al.}{2016}]{2016arXiv160406802K}
{Kashino} D.,  et~al., 2016, preprint, \href
  {http://adsabs.harvard.edu/abs/2016arXiv160406802K} {} (\mn@eprint {arXiv}
  {1604.06802})

\bibitem[\protect\citeauthoryear{{Kauffmann} et~al.,}{{Kauffmann}
  et~al.}{2003}]{2003MNRAS.346.1055K}
{Kauffmann} G.,  et~al., 2003, \mn@doi [\mnras]
  {10.1111/j.1365-2966.2003.07154.x}, \href
  {http://adsabs.harvard.edu/abs/2003MNRAS.346.1055K} {346, 1055}

\bibitem[\protect\citeauthoryear{{Kennicutt}}{{Kennicutt}}{1998}]{1998ApJ...498..541K}
{Kennicutt} Jr. R.~C.,  1998, \mn@doi [\apj] {10.1086/305588}, \href
  {http://adsabs.harvard.edu/abs/1998ApJ...498..541K} {498, 541}

\bibitem[\protect\citeauthoryear{{Kennicutt} \& {Evans}}{{Kennicutt} \&
  {Evans}}{2012}]{2012ARA&amp;A..50..531K}
{Kennicutt} R.~C.,  {Evans} N.~J.,  2012, \mn@doi [\araa]
  {10.1146/annurev-astro-081811-125610}, \href
  {http://adsabs.harvard.edu/abs/2012ARA%26A..50..531K} {50, 531}

\bibitem[\protect\citeauthoryear{{Kewley} \& {Dopita}}{{Kewley} \&
  {Dopita}}{2002}]{2002ApJS..142...35K}
{Kewley} L.~J.,  {Dopita} M.~A.,  2002, \mn@doi [\apjsupp] {10.1086/341326},
  \href {http://adsabs.harvard.edu/abs/2002ApJS..142...35K} {142, 35}

\bibitem[\protect\citeauthoryear{{Kewley}, {Dopita}, {Sutherland}, {Heisler}
  \& {Trevena}}{{Kewley} et~al.}{2001}]{2001ApJ...556..121K}
{Kewley} L.~J.,  {Dopita} M.~A.,  {Sutherland} R.~S.,  {Heisler} C.~A.,
  {Trevena} J.,  2001, \mn@doi [\apj] {10.1086/321545}, \href
  {http://adsabs.harvard.edu/abs/2001ApJ...556..121K} {556, 121}

\bibitem[\protect\citeauthoryear{{Kewley}, {Jansen}  \& {Geller}}{{Kewley}
  et~al.}{2005}]{2005PASP..117..227K}
{Kewley} L.~J.,  {Jansen} R.~A.,   {Geller} M.~J.,  2005, \mn@doi [\pasp]
  {10.1086/428303}, \href {http://adsabs.harvard.edu/abs/2005PASP..117..227K}
  {117, 227}

\bibitem[\protect\citeauthoryear{Kewley, Dopita, Leitherer, Dav{\'e}, Yuan,
  Allen, Groves  \& Sutherland}{Kewley et~al.}{2013}]{0004-637X-774-2-100}
Kewley L.~J.,  Dopita M.~A.,  Leitherer C.,  Dav{\'e} R.,  Yuan T.,  Allen M.,
  Groves B.,   Sutherland R.,  2013, \apj, 774, 100

\bibitem[\protect\citeauthoryear{{Kewley}, {Zahid}, {Geller}, {Dopita}, {Hwang}
   \& {Fabricant}}{{Kewley} et~al.}{2015}]{2015ApJ...812L..20K}
{Kewley} L.~J.,  {Zahid} H.~J.,  {Geller} M.~J.,  {Dopita} M.~A.,  {Hwang}
  H.~S.,   {Fabricant} D.,  2015, \mn@doi [\apjl]
  {10.1088/2041-8205/812/2/L20}, \href
  {http://adsabs.harvard.edu/abs/2015ApJ...812L..20K} {812, L20}

\bibitem[\protect\citeauthoryear{{Kobulnicky} \& {Kewley}}{{Kobulnicky} \&
  {Kewley}}{2004}]{2004ApJ...617..240K}
{Kobulnicky} H.~A.,  {Kewley} L.~J.,  2004, \mn@doi [\apj] {10.1086/425299},
  \href {http://adsabs.harvard.edu/abs/2004ApJ...617..240K} {617, 240}

\bibitem[\protect\citeauthoryear{{Krumholz} \& {Matzner}}{{Krumholz} \&
  {Matzner}}{2009}]{2009ApJ...703.1352K}
{Krumholz} M.~R.,  {Matzner} C.~D.,  2009, \mn@doi [\apj]
  {10.1088/0004-637X/703/2/1352}, \href
  {http://adsabs.harvard.edu/abs/2009ApJ...703.1352K} {703, 1352}

\bibitem[\protect\citeauthoryear{{Laigle} et~al.,}{{Laigle}
  et~al.}{2016}]{2016arXiv160402350L}
{Laigle} C.,  et~al., 2016, preprint, \href
  {http://adsabs.harvard.edu/abs/2016arXiv160402350L} {} (\mn@eprint {arXiv}
  {1604.02350})

\bibitem[\protect\citeauthoryear{Levesque, Kewley  \& Larson}{Levesque
  et~al.}{2010}]{1538-3881-139-2-712}
Levesque E.~M.,  Kewley L.~J.,   Larson K.~L.,  2010, The Astronomical Journal,
  139, 712

\bibitem[\protect\citeauthoryear{{Liu}, {Shapley}, {Coil}, {Brinchmann}  \&
  {Ma}}{{Liu} et~al.}{2008}]{2008ApJ...678..758L}
{Liu} X.,  {Shapley} A.~E.,  {Coil} A.~L.,  {Brinchmann} J.,   {Ma} C.-P.,
  2008, \mn@doi [\apj] {10.1086/529030}, \href
  {http://adsabs.harvard.edu/abs/2008ApJ...678..758L} {678, 758}

\bibitem[\protect\citeauthoryear{{Madau} \& {Dickinson}}{{Madau} \&
  {Dickinson}}{2014}]{2014ARA&amp;A..52..415M}
{Madau} P.,  {Dickinson} M.,  2014, \mn@doi [\araa]
  {10.1146/annurev-astro-081811-125615}, \href
  {http://adsabs.harvard.edu/abs/2014ARA%26A..52..415M} {52, 415}

\bibitem[\protect\citeauthoryear{{Masters} et~al.,}{{Masters}
  et~al.}{2014}]{2014ApJ...785..153M}
{Masters} D.,  et~al., 2014, \mn@doi [\apj] {10.1088/0004-637X/785/2/153},
  \href {http://adsabs.harvard.edu/abs/2014ApJ...785..153M} {785, 153}

\bibitem[\protect\citeauthoryear{{McCracken} et~al.,}{{McCracken}
  et~al.}{2012}]{2012A&amp;A...544A.156M}
{McCracken} H.~J.,  et~al., 2012, \mn@doi [\aap] {10.1051/0004-6361/201219507},
  \href {http://adsabs.harvard.edu/abs/2012A%26A...544A.156M} {544, A156}

\bibitem[\protect\citeauthoryear{{Mesa-Delgado}, {N{\'u}{\~n}ez-D{\'{\i}}az},
  {Esteban}, {L{\'o}pez-Mart{\'{\i}}n}  \&
  {Garc{\'{\i}}a-Rojas}}{{Mesa-Delgado} et~al.}{2011}]{2011MNRAS.417..420M}
{Mesa-Delgado} A.,  {N{\'u}{\~n}ez-D{\'{\i}}az} M.,  {Esteban} C.,
  {L{\'o}pez-Mart{\'{\i}}n} L.,   {Garc{\'{\i}}a-Rojas} J.,  2011, \mn@doi
  [\mnras] {10.1111/j.1365-2966.2011.19278.x}, \href
  {http://adsabs.harvard.edu/abs/2011MNRAS.417..420M} {417, 420}

\bibitem[\protect\citeauthoryear{{Murphy} et~al.,}{{Murphy}
  et~al.}{2011}]{2011ApJ...737...67M}
{Murphy} E.~J.,  et~al., 2011, \mn@doi [\apj] {10.1088/0004-637X/737/2/67},
  \href {http://adsabs.harvard.edu/abs/2011ApJ...737...67M} {737, 67}

\bibitem[\protect\citeauthoryear{{Nakajima} \& {Ouchi}}{{Nakajima} \&
  {Ouchi}}{2014}]{2014MNRAS.442..900N}
{Nakajima} K.,  {Ouchi} M.,  2014, \mn@doi [\mnras] {10.1093/mnras/stu902},
  \href {http://adsabs.harvard.edu/abs/2014MNRAS.442..900N} {442, 900}

\bibitem[\protect\citeauthoryear{{Newman} et~al.,}{{Newman}
  et~al.}{2013}]{2013ApJS..208....5N}
{Newman} J.~A.,  et~al., 2013, \mn@doi [\apjs] {10.1088/0067-0049/208/1/5},
  \href {http://adsabs.harvard.edu/abs/2013ApJS..208....5N} {208, 5}

\bibitem[\protect\citeauthoryear{{Newman} et~al.,}{{Newman}
  et~al.}{2014}]{2014ApJ...781...21N}
{Newman} S.~F.,  et~al., 2014, \mn@doi [\apj] {10.1088/0004-637X/781/1/21},
  \href {http://adsabs.harvard.edu/abs/2014ApJ...781...21N} {781, 21}

\bibitem[\protect\citeauthoryear{{Oke}}{{Oke}}{1990}]{1990AJ.....99.1621O}
{Oke} J.~B.,  1990, \mn@doi [\aj] {10.1086/115444}, \href
  {http://adsabs.harvard.edu/abs/1990AJ.....99.1621O} {99, 1621}

\bibitem[\protect\citeauthoryear{{Osterbrock} \& {Ferland}}{{Osterbrock} \&
  {Ferland}}{2006}]{2006agna.book.....O}
{Osterbrock} D.~E.,  {Ferland} G.~J.,  2006, {Astrophysics of gaseous nebulae
  and active galactic nuclei}

\bibitem[\protect\citeauthoryear{{Rigby}, {Wuyts}, {Gladders}, {Sharon}  \&
  {Becker}}{{Rigby} et~al.}{2011}]{2011ApJ...732...59R}
{Rigby} J.~R.,  {Wuyts} E.,  {Gladders} M.~D.,  {Sharon} K.,   {Becker} G.~D.,
  2011, \mn@doi [\apj] {10.1088/0004-637X/732/1/59}, \href
  {http://adsabs.harvard.edu/abs/2011ApJ...732...59R} {732, 59}

\bibitem[\protect\citeauthoryear{{Rudnick} et~al.,}{{Rudnick}
  et~al.}{2003}]{2003ApJ...599..847R}
{Rudnick} G.,  et~al., 2003, \mn@doi [\apj] {10.1086/379628}, \href
  {http://adsabs.harvard.edu/abs/2003ApJ...599..847R} {599, 847}

\bibitem[\protect\citeauthoryear{{Sanders} et~al.,}{{Sanders}
  et~al.}{2016}]{2016ApJ...816...23S}
{Sanders} R.~L.,  et~al., 2016, \mn@doi [\apj] {10.3847/0004-637X/816/1/23},
  \href {http://adsabs.harvard.edu/abs/2016ApJ...816...23S} {816, 23}

\bibitem[\protect\citeauthoryear{{Scoville} et~al.,}{{Scoville}
  et~al.}{2007}]{2007ApJS..172....1S}
{Scoville} N.,  et~al., 2007, \mn@doi [\apjs] {10.1086/516585}, \href
  {http://adsabs.harvard.edu/abs/2007ApJS..172....1S} {172, 1}

\bibitem[\protect\citeauthoryear{Shapley et~al.,}{Shapley
  et~al.}{2015}]{0004-637X-801-2-88}
Shapley A.~E.,  et~al., 2015, \apj, 801, 88

\bibitem[\protect\citeauthoryear{{Shimakawa} et~al.,}{{Shimakawa}
  et~al.}{2015}]{2015MNRAS.451.1284S}
{Shimakawa} R.,  et~al., 2015, \mn@doi [\mnras] {10.1093/mnras/stv915}, \href
  {http://adsabs.harvard.edu/abs/2015MNRAS.451.1284S} {451, 1284}

\bibitem[\protect\citeauthoryear{{Shirazi}, {Vegetti}, {Nesvadba}, {Allam},
  {Brinchmann}  \& {Tucker}}{{Shirazi} et~al.}{2014a}]{2014MNRAS.440.2201S}
{Shirazi} M.,  {Vegetti} S.,  {Nesvadba} N.,  {Allam} S.,  {Brinchmann} J.,
  {Tucker} D.,  2014a, \mn@doi [\mnras] {10.1093/mnras/stu316}, \href
  {http://adsabs.harvard.edu/abs/2014MNRAS.440.2201S} {440, 2201}

\bibitem[\protect\citeauthoryear{{Shirazi}, {Brinchmann}  \&
  {Rahmati}}{{Shirazi} et~al.}{2014b}]{2014ApJ...787..120S}
{Shirazi} M.,  {Brinchmann} J.,   {Rahmati} A.,  2014b, \mn@doi [\apj]
  {10.1088/0004-637X/787/2/120}, \href
  {http://adsabs.harvard.edu/abs/2014ApJ...787..120S} {787, 120}

\bibitem[\protect\citeauthoryear{{Silverman} et~al.,}{{Silverman}
  et~al.}{2014}]{2014arXiv1409.0447S}
{Silverman} J.~D.,  et~al., 2014, preprint, \href
  {http://adsabs.harvard.edu/abs/2014arXiv1409.0447S} {} (\mn@eprint {arXiv}
  {1409.0447})

\bibitem[\protect\citeauthoryear{{Speagle}, {Steinhardt}, {Capak}  \&
  {Silverman}}{{Speagle} et~al.}{2014}]{2014ApJS..214...15S}
{Speagle} J.~S.,  {Steinhardt} C.~L.,  {Capak} P.~L.,   {Silverman} J.~D.,
  2014, \mn@doi [\apjs] {10.1088/0067-0049/214/2/15}, \href
  {http://adsabs.harvard.edu/abs/2014ApJS..214...15S} {214, 15}

\bibitem[\protect\citeauthoryear{{Steidel}, {Shapley}, {Pettini}, {Adelberger},
  {Erb}, {Reddy}  \& {Hunt}}{{Steidel} et~al.}{2004}]{2004ApJ...604..534S}
{Steidel} C.~C.,  {Shapley} A.~E.,  {Pettini} M.,  {Adelberger} K.~L.,  {Erb}
  D.~K.,  {Reddy} N.~A.,   {Hunt} M.~P.,  2004, \mn@doi [\apj]
  {10.1086/381960}, \href {http://adsabs.harvard.edu/abs/2004ApJ...604..534S}
  {604, 534}

\bibitem[\protect\citeauthoryear{{Steidel} et~al.,}{{Steidel}
  et~al.}{2014}]{2014ApJ...795..165S}
{Steidel} C.~C.,  et~al., 2014, \mn@doi [\apj] {10.1088/0004-637X/795/2/165},
  \href {http://adsabs.harvard.edu/abs/2014ApJ...795..165S} {795, 165}

\bibitem[\protect\citeauthoryear{{Tacconi} et~al.,}{{Tacconi}
  et~al.}{2010}]{2010Natur.463..781T}
{Tacconi} L.~J.,  et~al., 2010, \mn@doi [\nat] {10.1038/nature08773}, \href
  {http://adsabs.harvard.edu/abs/2010Natur.463..781T} {463, 781}

\bibitem[\protect\citeauthoryear{{Tayal}}{{Tayal}}{2007}]{2007ApJS..171..331T}
{Tayal} S.~S.,  2007, \mn@doi [\apjs] {10.1086/513107}, \href
  {http://adsabs.harvard.edu/abs/2007ApJS..171..331T} {171, 331}

\bibitem[\protect\citeauthoryear{{Tayal} \& {Zatsarinny}}{{Tayal} \&
  {Zatsarinny}}{2010}]{2010ApJS..188...32T}
{Tayal} S.~S.,  {Zatsarinny} O.,  2010, \mn@doi [\apjs]
  {10.1088/0067-0049/188/1/32}, \href
  {http://adsabs.harvard.edu/abs/2010ApJS..188...32T} {188, 32}

\bibitem[\protect\citeauthoryear{{Tremonti} et~al.,}{{Tremonti}
  et~al.}{2004}]{2004ApJ...613..898T}
{Tremonti} C.~A.,  et~al., 2004, \mn@doi [\apj] {10.1086/423264}, \href
  {http://adsabs.harvard.edu/abs/2004ApJ...613..898T} {613, 898}

\bibitem[\protect\citeauthoryear{{Veilleux} \& {Osterbrock}}{{Veilleux} \&
  {Osterbrock}}{1987}]{1987ApJS...63..295V}
{Veilleux} S.,  {Osterbrock} D.~E.,  1987, \mn@doi [\apjsupp] {10.1086/191166},
  \href {http://adsabs.harvard.edu/abs/1987ApJS...63..295V} {63, 295}

\bibitem[\protect\citeauthoryear{{York} et~al.,}{{York}
  et~al.}{2000}]{2000AJ....120.1579Y}
{York} D.~G.,  et~al., 2000, \mn@doi [\aj] {10.1086/301513}, \href
  {http://adsabs.harvard.edu/abs/2000AJ....120.1579Y} {120, 1579}

\bibitem[\protect\citeauthoryear{{Zahid}, {Dima}, {Kewley}, {Erb}  \&
  {Dav{\'e}}}{{Zahid} et~al.}{2012}]{2012ApJ...757...54Z}
{Zahid} H.~J.,  {Dima} G.~I.,  {Kewley} L.~J.,  {Erb} D.~K.,   {Dav{\'e}} R.,
  2012, \mn@doi [\apj] {10.1088/0004-637X/757/1/54}, \href
  {http://adsabs.harvard.edu/abs/2012ApJ...757...54Z} {757, 54}

\bibitem[\protect\citeauthoryear{Zahid et~al.,}{Zahid
  et~al.}{2014}]{0004-637X-792-1-75}
Zahid H.~J.,  et~al., 2014, \apj, 792, 75

\bibitem[\protect\citeauthoryear{{van der Wel} et~al.,}{{van der Wel}
  et~al.}{2014}]{2014ApJ...788...28V}
{van der Wel} A.,  et~al., 2014, \mn@doi [\apj] {10.1088/0004-637X/788/1/28},
  \href {http://adsabs.harvard.edu/abs/2014ApJ...788...28V} {788, 28}

\makeatother
\end{thebibliography}

% \begin{thebibliography}{99}
% \bibitem[\protect\citeauthoryear{Author}{2012}]{Author2012}
% Author A.~N., 2013, Journal of Improbable Astronomy, 1, 1
% \bibitem[\protect\citeauthoryear{Others}{2013}]{Others2013}
% Others S., 2012, Journal of Interesting Stuff, 17, 198
% \end{thebibliography}

%%%%%%%%%%%%%%%%%%%%%%%%%%%%%%%%%%%%%%%%%%%%%%%%%%

%%%%%%%%%%%%%%%%% APPENDICES %%%%%%%%%%%%%%%%%%%%%

\appendix

\section{Data}

% \tiny{
% \begin{landscape}
\begin{table*}
\caption{Summary of measurements for the $z\sim 1.5$ \OII-\Ha\ detected sample}
\label{tab:main_sample}
\ra{1.2}
\begin{tabular}{@{}lcccccccccc@{}}
	\toprule
 	DEIMOS     					&
 	$\alpha$                      &
 	$\delta$                      &
 	z mag                        	&
 	$\mathrm{z_{spec}}$   		&
 	$\log \Mstar$                 &
 	$\log \SFR$                   &
 	$\OII$                       &
 	$\OII\lambda 3726$            &
 	$\OII \lambda 3729$           &
 	$n_e$                         \\ %&
 	% OH flag                      \\
 	% units for each column
 	ID                           	&
 	(hr)                         	&
 	(deg)                         &
 	                           	&
 	\OII             				&
 	[$\Msun$]                  	&
 	[$\Msunyr$]                  	&
 	\multicolumn{3}{c}{[$10^{-17}\flux$]}  &
 	[\cmcub]                     \\
 	%column number
 	(1)                        &
 	(2)                        &
 	(3)                        &
 	(4)                        &
 	(5)                        &
 	(6)                        &
 	(7)                        &
 	(8)                        &
 	(9)                        &
 	(10)                       &
 	(11)                       \\ %&
 	% (12)                       \\
 	\toprule
% start data
JK\_28823  & 09:59:56.98 & +02:09:20.5 & 22.1  & $1.406$ & 10.95 & $2.31\pm0.22$ & $2.99\pm0.27$ & $2.60\pm0.20^*$ & $0.40\pm0.18^*$ &   \\ %            1  \\  
D345075 & 09:59:37.82 & +02:14:23.2 & 23.4  & $1.427$ & 10.66 & $1.11\pm0.03$ & $4.09\pm0.30$ & $1.65\pm0.17^*$ & $2.44\pm0.23^*$ &  \\ %             1  \\  
D464854 & 09:59:25.37 & +02:30:47.0 & 22.6  & $1.434$ & 10.59 & $1.30\pm0.05$ & $6.21\pm0.20$ & $2.59\pm0.12\ $ & $3.62\pm0.12\ $ & $35^{+51}_{-35}$  \\ %            0  \\  
D409473 & 10:00:47.61 & +02:23:27.5 & 22.7  & $1.436$ & 10.48 & $1.20\pm0.05$ & $3.36\pm0.14$ & $1.40\pm0.08\ $ & $1.96\pm0.10\ $ & $34^{+72}_{-34}$  \\ %            0  \\  
JK\_38652  & 10:00:00.63 & +02:33:01.2 & 22.6  & $1.441$ & 11.87 & $1.88\pm0.24$ & $2.14\pm0.40$ & $1.04\pm0.80^*$ & $1.10\pm0.81^*$ &  \\ %            1  \\  
D341519 & 10:00:29.08 & +02:13:55.2 & 23.6  & $1.444$ & 10.29 & $0.75\pm0.05$ & $1.60\pm0.08$ & $0.65\pm0.05\ $ & $0.94\pm0.05\ $ & $9^{+75}_{-9}$ \\ %            0  \\  
D387090 & 09:59:19.37 & +02:20:14.0 & 22.6  & $1.452$ & 10.68 & $1.39\pm0.26$ & $3.98\pm0.20$ & $1.82\pm0.16\ $ & $2.16\pm0.18\ $ & $218^{+216}_{-143}$ \\ %             0  \\  
D455626 & 09:59:55.34 & +02:29:33.5 & 22.9  & $1.452$ & 10.77 & $1.45\pm0.05$ & $4.62\pm0.27$ & $1.75\pm0.16\ $ & $2.87\pm0.18\ $ & $<10$ \\ %             0  \\  
1038215 & 09:59:56.32 & +02:10:03.0 & 23.3  & $1.457$ & 9.64  & $1.13\pm0.05$ & $3.36\pm0.22$ & $1.46\pm0.16\ $ & $1.90\pm0.14\ $ & $113^{+171}_{-113}$ \\ %            0  \\  
D298530 & 09:59:42.10 & +02:08:06.3 & 22.2  & $1.470$ & 11.54 & $1.71\pm0.18$ & $0.50\pm0.13$ & $0.25\pm0.08\ $ & $0.25\pm0.09\ $ &   \\ %             0  \\  
D451530 & 09:59:16.42 & +02:29:02.3 & 22.7  & $1.471$ & 10.90 & $1.70\pm0.05$ & $9.05\pm0.25$ & $4.50\pm0.18\ $ & $4.55\pm0.19\ $ & $451^{+111}_{-92}$  \\ %            0  \\  
D344598 & 09:59:28.09 & +02:14:14.7 & 23.0  & $1.475$ & 11.18 & $1.25\pm0.05$ & $1.91\pm0.17$ & $1.07\pm0.14\ $ & $0.85\pm0.12\ $ & $1018^{+1106}_{-473}$ \\ %             0  \\
G174035 & 09:59:28.92 & +02:14:34.3 & 22.6  & $1.475$ & 11.10 & $1.77\pm0.05$ & $3.12\pm0.41$ & $1.00\pm0.76\ $ & $2.12\pm0.99\ $ &   \\ %            0  \\  
G-9848  & 09:59:32.26 & +02:04:09.6 & 23.3  & $1.487$ & 10.53 & $1.26\pm0.11$ & $1.38\pm0.12$ & $0.55\pm0.07\ $ & $0.83\pm0.08\ $ & $<10$ \\ %            0  \\  
G1190 & 09:59:47.15 & +02:06:27.5 & 22.6  & $1.489$ & 11.09 & $1.53\pm0.05$ & $1.86\pm0.16$ & $0.62\pm0.12\ $ & $1.24\pm0.12\ $ & $<10$ \\ %            0  \\  
D394728 & 09:59:26.10 & +02:21:20.8 & 22.3  & $1.501$ & 10.72 & $1.14\pm0.07$ & $11.12\pm0.54$  & $5.12\pm0.53^*$ & $6.00\pm0.27^*$ &   \\ %            1  \\  
G158505 & 09:59:14.52 & +02:11:33.9 & 23.1  & $1.504$ & 10.86 & $1.64\pm0.05$ & $4.16\pm0.85$ & $1.93\pm0.49^*$ & $2.23\pm0.67^*$ &  \\ %            1  \\  
D319520 & 09:59:47.56 & +02:10:52.7 & 22.6  & $1.505$ & 10.82 & $0.67\pm0.13$ & $6.00\pm0.69$ & $1.85\pm3.15^*$ & $4.15\pm3.66^*$ &   \\ %            1  \\  
774526  & 10:00:20.96 & +02:04:07.4 & 23.9  & $1.506$ & 10.38 & $0.70\pm0.18$ & $2.48\pm0.19$ & $1.20\pm0.13\ $ & $1.28\pm0.11\ $ & $365^{+245}_{-164}$ \\ %            0  \\  
G11751  & 09:59:40.07 & +02:08:31.7 & 23.2  & $1.507$ & 10.28 & $1.28\pm0.05$ & $4.38\pm0.32$ & $1.73\pm0.17\ $ & $2.65\pm0.22\ $ & $<10$ \\ %            0  \\ 
JK\_16807  & 10:00:20.59 & +02:17:07.2 & 23.1  & $1.515$ & 11.11 & $0.59\pm0.16$ & $2.34\pm0.41$ & $1.34\pm0.41^*$ & $1.00\pm0.44^*$ &   \\ %            1  \\  
797988  & 09:59:50.77 & +02:04:49.9 & 22.8  & $1.519$ & 11.16 & $1.26\pm0.24$ & $2.63\pm0.53$ & $1.57\pm0.51^*$ & $1.06\pm0.21^*$ &   \\ %            1  \\  
D490890 & 09:59:28.30 & +02:34:18.0 & 22.6  & $1.522$ & 11.24 & $1.05\pm0.07$ & $1.57\pm0.90$ & $0.83\pm0.77\ $ & $0.74\pm0.57\ $ &   \\ %            0  \\  
G-14013 & 10:00:23.15 & +02:03:18.0 & 23.6  & $1.524$ & 10.77 & $0.67\pm0.05$ & $2.59\pm0.33$ & $1.19\pm0.26\ $ & $1.41\pm0.22\ $ &   \\ %            0  \\  
D308265 & 09:59:27.44 & +02:09:27.7 & 23.1  & $1.524$ & 10.57 & $1.17\pm0.05$ & $4.29\pm0.17$ & $1.73\pm0.10\ $ & $2.57\pm0.11\ $ & $<10$ \\ %            0  \\  
D325472 & 09:59:21.14 & +02:11:40.8 & 22.9  & $1.526$ & 10.55 & $1.04\pm0.18$ & $7.93\pm0.51$ & $3.11\pm0.21^*$ & $4.81\pm0.48^*$ &   \\ %            1  \\  
D318267 & 09:59:18.30 & +02:10:44.8 & 23.3  & $1.526$ & 10.20 & $1.15\pm0.05$ & $5.62\pm0.54$ & $2.20\pm0.27^*$ & $3.42\pm0.57^*$ &   \\ %            1  \\  
1068560 & 09:59:34.50 & +02:07:46.6 & 24.0  & $1.540$ & 10.06 & $0.85\pm0.05$ & $2.00\pm1.10$ & $2.00\pm1.06^*$ & $0.00\pm0.43^*$ &   \\ %            1  \\  
JK\_41000  & 09:59:38.94 & +02:16:53.9 & 22.2  & $1.551$ & 11.60 & $1.84\pm0.05$ & $4.15\pm0.73$ & $3.39\pm1.22\ $ & $0.76\pm1.23\ $ &   \\ %&            0  \\
D339009 & 09:59:15.01 & +02:13:33.8 & 23.2  & $1.550$ & 9.96  & $1.16\pm0.05$ & $3.69\pm0.38$ & $1.95\pm0.37^*$ & $1.74\pm0.20^*$ &   \\ %&            1  \\  
G26349  & 09:59:42.09 & +02:11:23.2 & 22.7  & $1.551$ & 11.36 & $1.65\pm0.05$ & $5.86\pm1.11$ & $2.50\pm2.99^*$ & $3.36\pm3.65^*$ &   \\ %&            1  \\  
D340558 & 09:59:38.19 & +02:13:40.7 & 23.1  & $1.580$ & 10.40 & $1.03\pm0.05$ & $5.66\pm0.59$ & $1.88\pm0.29^*$ & $3.78\pm0.52^*$ &   \\ %&            1  \\  
D307756 & 09:59:42.34 & +02:09:21.9 & 23.3  & $1.583$ & 10.90 & $0.44\pm0.05$ & $6.82\pm0.28$ & $2.97\pm0.23\ $ & $3.86\pm0.25\ $ & $108^{+144}_{-104}$ \\ %&         0  \\  
D416912 & 10:00:40.61 & +02:24:28.0 & 22.8  & $1.587$ & 10.65 & $1.37\pm0.20$ & $5.44\pm0.39$ & $2.40\pm0.26\ $ & $3.04\pm0.29\ $ & $140^{+208}_{-135}$ \\ %&         0  \\  
D332067 & 09:59:24.05 & +02:12:35.7 & 22.6  & $1.587$ & 10.40 & $1.29\pm0.14$ & $12.23\pm0.38$  & $5.36\pm0.25\ $ & $6.87\pm0.23\ $ & $124^{+61}_{-52}$ \\ %&         0  \\  
220419\_HJZ  & 10:00:46.57 & +02:23:35.7 & 22.7  & $1.588$ & 10.76 & $1.22\pm0.05$ & $5.10\pm0.33$ & $2.05\pm0.23\ $ & $3.05\pm0.26\ $ & $<10$ \\ %&         0   \\  
D358016 & 09:59:22.34 & +02:16:10.7 & 23.3  & $1.587$ & 10.48 & $1.17\pm0.18$ & $3.19\pm0.26$ & $1.39\pm0.16\ $ & $1.80\pm0.18\ $ & $113^{+180}_{-113}$ \\ %&         0  \\  
G-8423  & 09:59:50.59 & +02:04:26.0 & 23.0  & $1.595$ & 10.98 & $0.82\pm0.05$ & $1.46\pm0.24$ & $0.55\pm0.17^*$ & $0.91\pm0.18^*$ &   \\ %&            1  \\  
D399476 & 09:59:29.23 & +02:22:01.0 & 23.5  & $1.617$ & 10.47 & $1.05\pm0.05$ & $5.14\pm0.56$ & $0.80\pm1.20\ $ & $4.34\pm1.30\ $ &   \\ %&            0  \\  
D352264 & 09:59:46.98 & +02:15:20.4 & 22.7  & $1.636$ & 10.68 & $1.71\pm0.05$ & $5.04\pm0.37$ & $2.21\pm0.24\ $ & $2.83\pm0.25\ $ & $125^{+178}_{-121}$ \\ %&            0  \\  
G-9661  & 09:59:49.26 & +02:04:09.7 & 23.3  & $1.638$ & 10.90 & $0.52\pm0.17$ & $2.66\pm0.38$ & $1.39\pm0.24\ $ & $1.27\pm0.23\ $ & $641^{+847}_{-364}$ \\ %&            0  \\  
G163773 & 10:00:22.60 & +02:12:34.2 & 23.3  & $1.641$ & 10.47 & $1.20\pm0.05$ & $4.89\pm0.81$ & $1.28\pm0.41^*$ & $3.61\pm0.75^*$ &   \\ %&            1  \\  
JK\_16428  & 09:59:41.31 & +02:14:42.8 & 23.9  & $1.647$ & 10.67 & $1.36\pm0.17$ & $8.43\pm0.63$ & $5.44\pm6.47^*$ & $2.99\pm6.69^*$ &   \\ %&            1  \\
G133455 & 09:59:43.00 & +02:06:36.7 & 23.5  & $1.652$ & 9.81  & $1.32\pm0.26$ & $9.30\pm0.60$ & $4.32\pm0.44\ $ & $4.98\pm0.43\ $ & $250^{+225}_{-149}$ \\ %&            0  \\  
1032970 & 09:59:48.39 & +02:12:09.2 & 22.6  & $1.654$ & 10.82 & $1.73\pm0.05$ & $9.66\pm0.84$* & $3.58\pm0.46^*$ & $6.08\pm0.64^*$ &   \\ %&            1  \\  
JK\_17606  & 10:00:36.31 & +02:21:17.5 & 23.0  & $1.654$ & 10.54 & $2.02\pm0.21$ & $7.55\pm0.94$ & $3.09\pm0.87^*$ & $4.45\pm0.83^*$ &   \\ %&            1  \\
%\tablenotetext{a}{Stellar masses from \cite{2016arXiv160402350L} are scaled to a Kroupa IMF.}
\bottomrule
\vspace{0.5ex}
\end{tabular}
\raggedright{ \textbf{Notes: }
 (1): DEIMOS identifier assigned to target 
 (2): Right ascenscion (J2000) in units of hours, minutes, and seconds. 
 (3): Declination (J2000) in units of degrees, arcminutes, and arcseconds
 (4): Z(AB) magnitude
 (5): Spectroscopic redshift determined from \OII\
 (6): $\log(\Mstar/\Msun)$ from BC03 best-fit template taken at the minimum $\chi^2$\citep{2016arXiv160402350L} 
 (7): $\log (\SFR / \Msunyr)$ Calculated from \Ha\ based on \cite{1989ApJ...345..245C} treatment of extinction 
 (8): Total \OII\llambda 3726,3729 flux 
 (9): $\OII\lambda 3726$ flux 
 (10): $\OII\lambda 3726$ flux 
 (11): Electron density estimated from $\OII\lambda 3726/\lambda 3729$
 % (12): Flag assigned to account for contamination by OH lines - a value of 1 is used to note the presence of OH lines in the region of the \OII\ doublet 
}
\end{table*}
% \end{landscape}
% }

\begin{figure*}
	\centering
		\begin{minipage}{0.98\linewidth} %trim={<left> <lower> <right> <upper>}
		\includegraphics[width=0.5\textwidth,trim={0cm 0cm 0.5cm 0cm},clip]{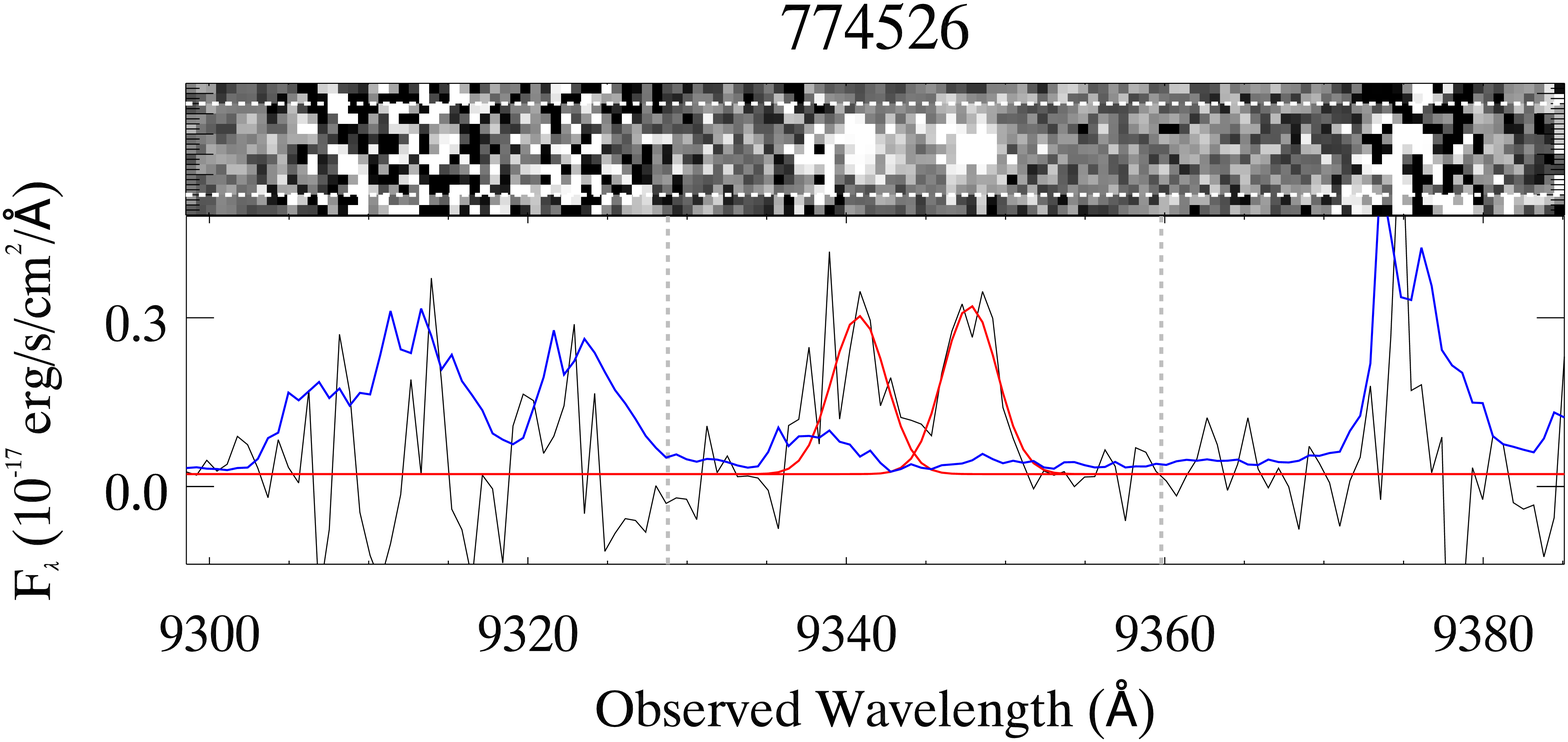}
		\includegraphics[width=0.5\textwidth,trim={0cm 0cm 0.5cm 0cm},clip]{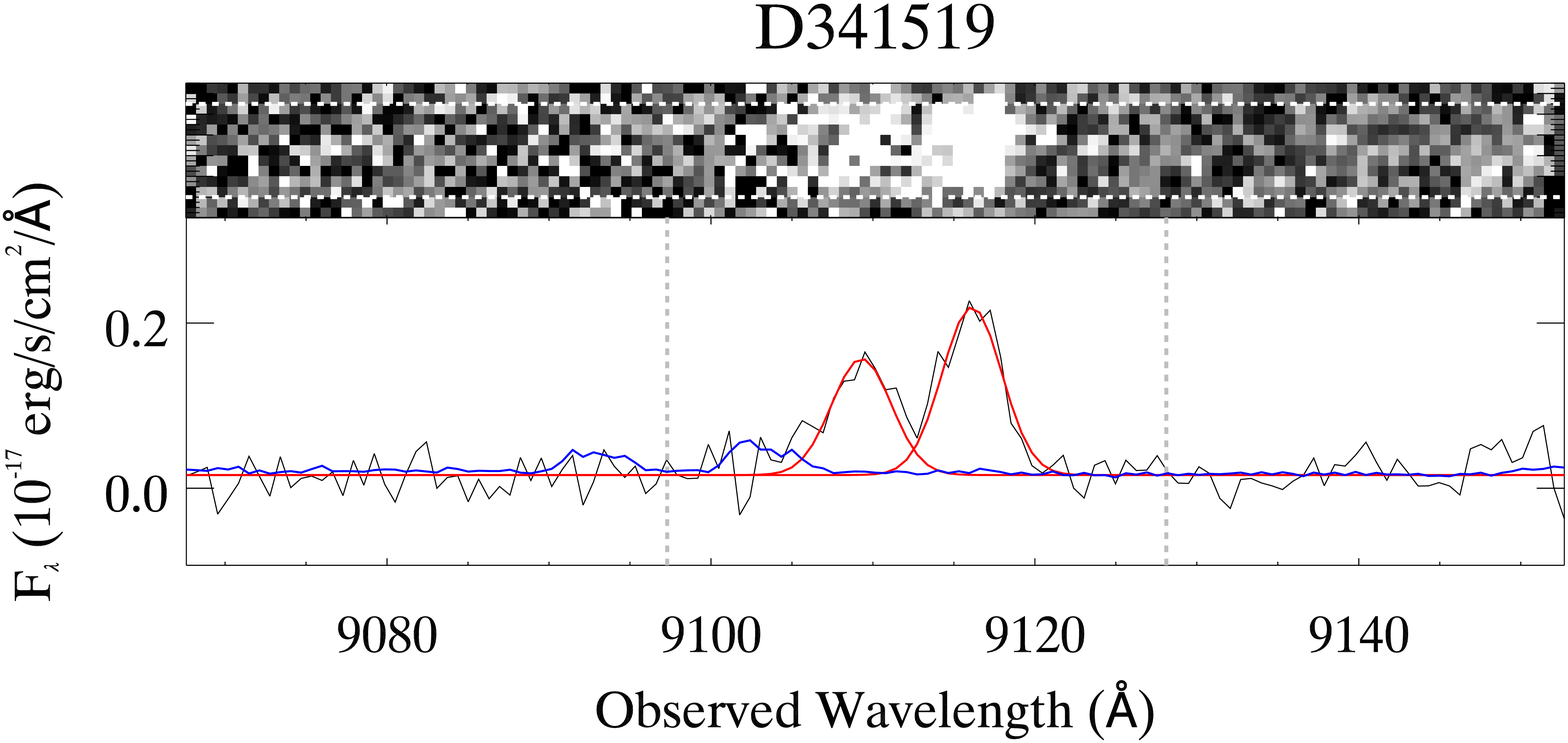}
		\includegraphics[width=0.5\textwidth,trim={0cm 0cm 0.5cm 0cm},clip]{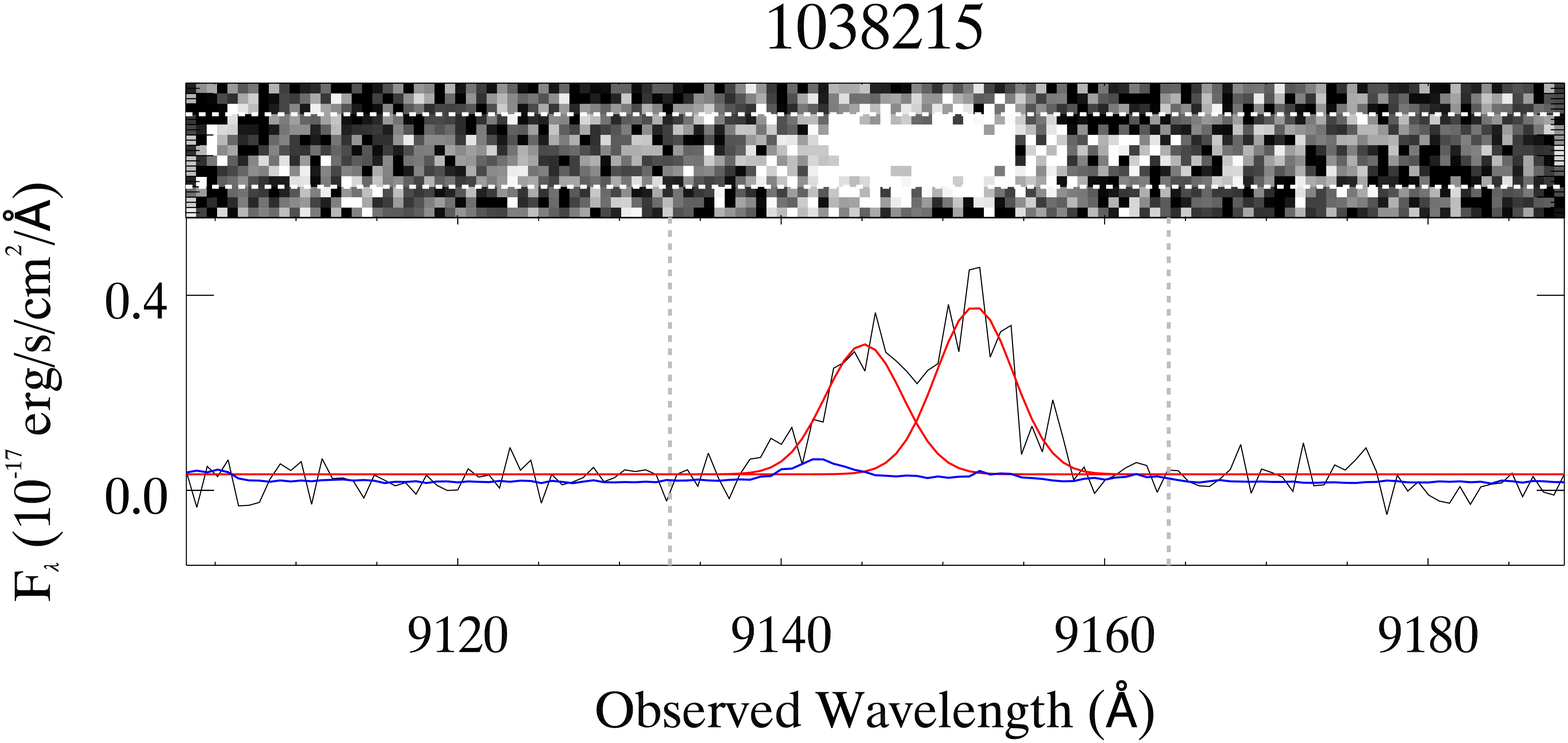}
		\includegraphics[width=0.5\textwidth,trim={0cm 0cm 0.5cm 0cm},clip]{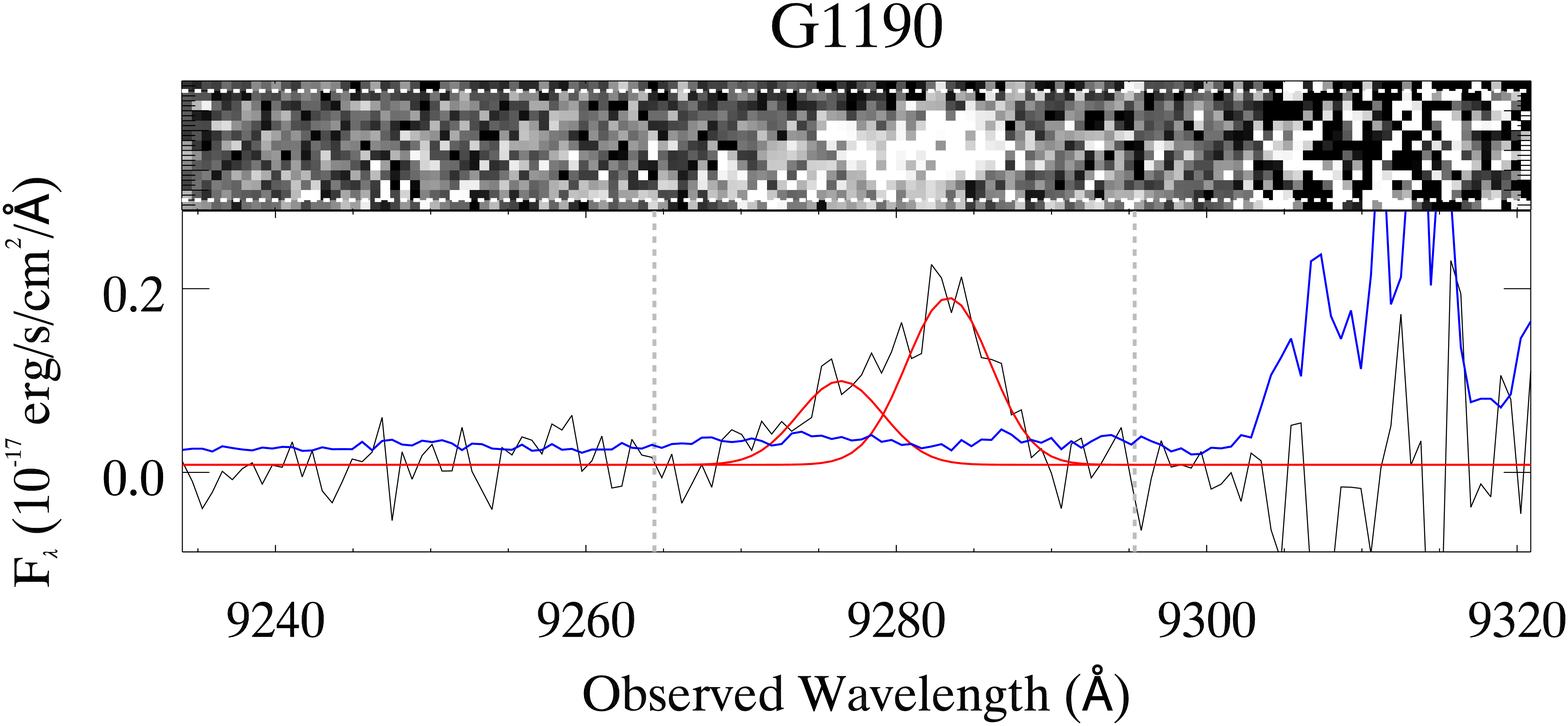}
		\includegraphics[width=0.5\textwidth,trim={0cm 0cm 0.5cm 0cm},clip]{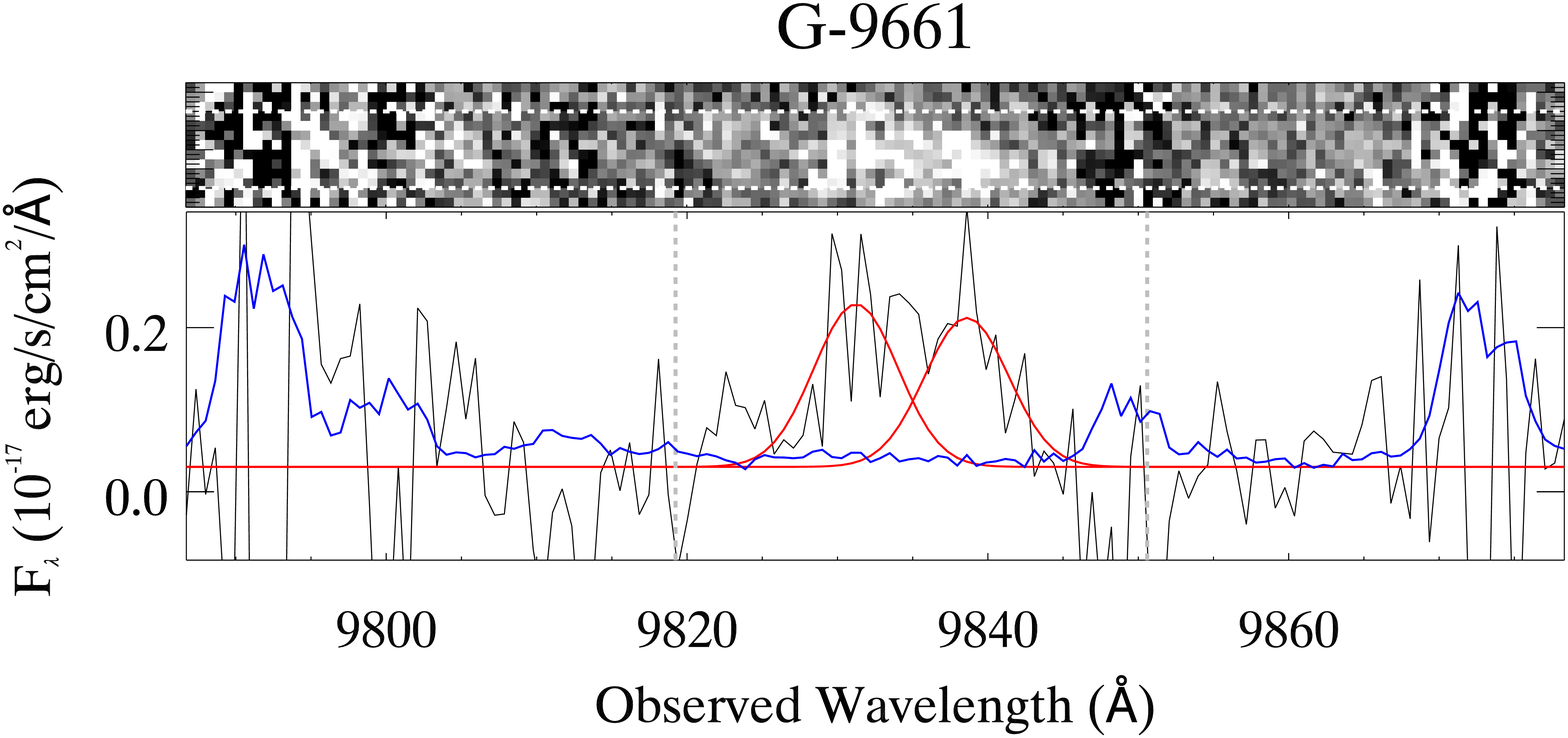}
		\includegraphics[width=0.5\textwidth,trim={0cm 0cm 0.5cm 0cm},clip]{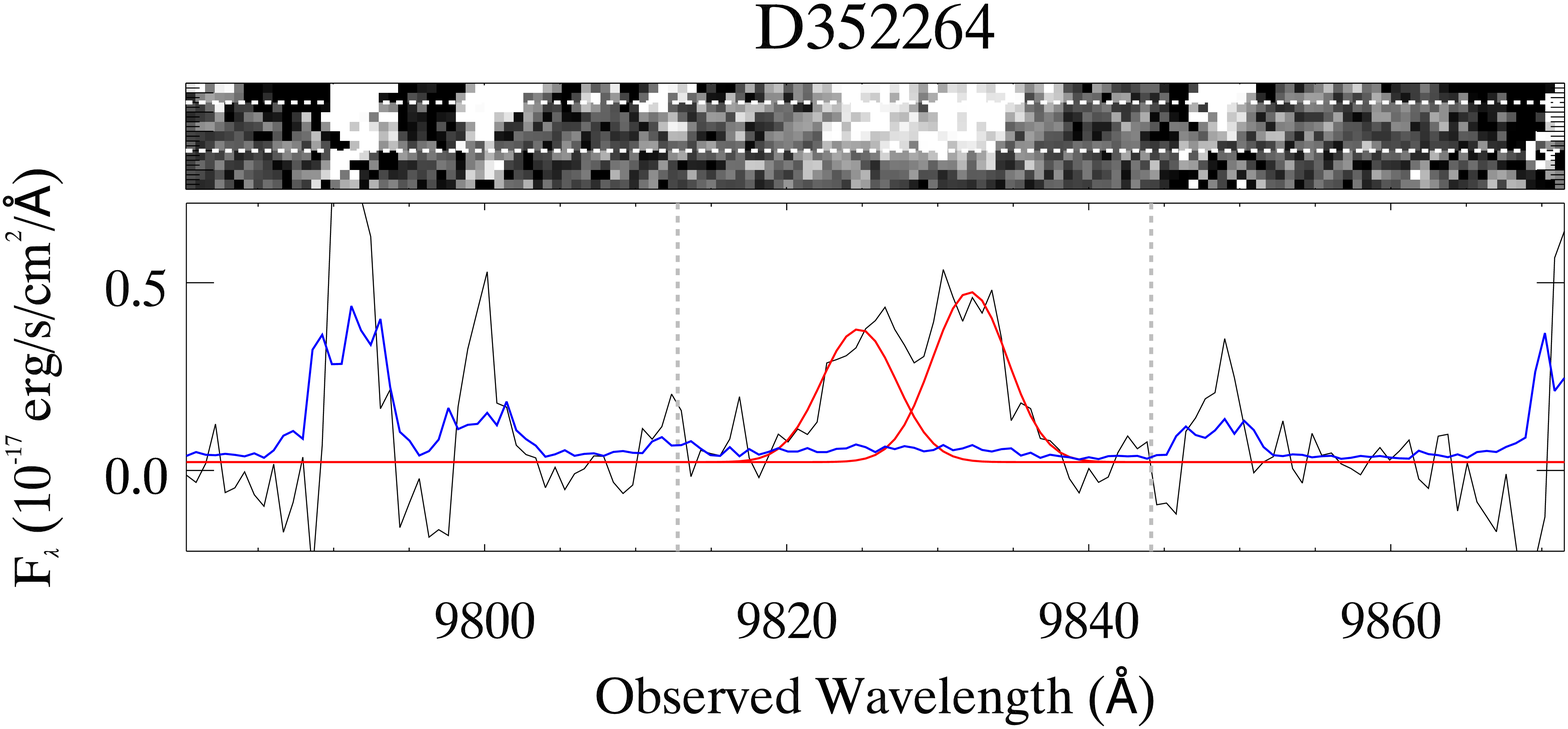}
		\includegraphics[width=0.5\textwidth,trim={0cm 0cm 0.5cm 0cm},clip]{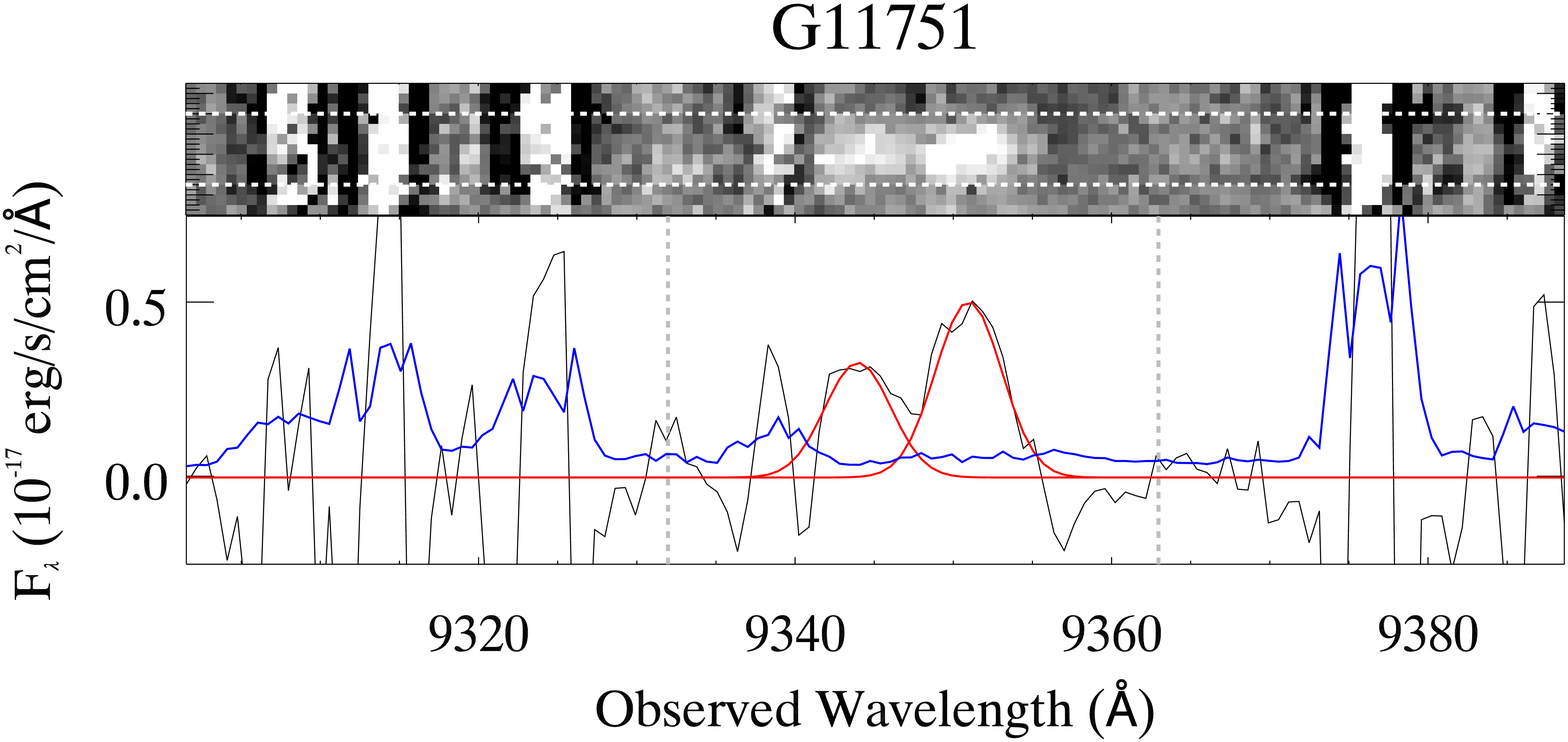}
		\includegraphics[width=0.5\textwidth,trim={0cm 0cm 0.5cm 0cm},clip]{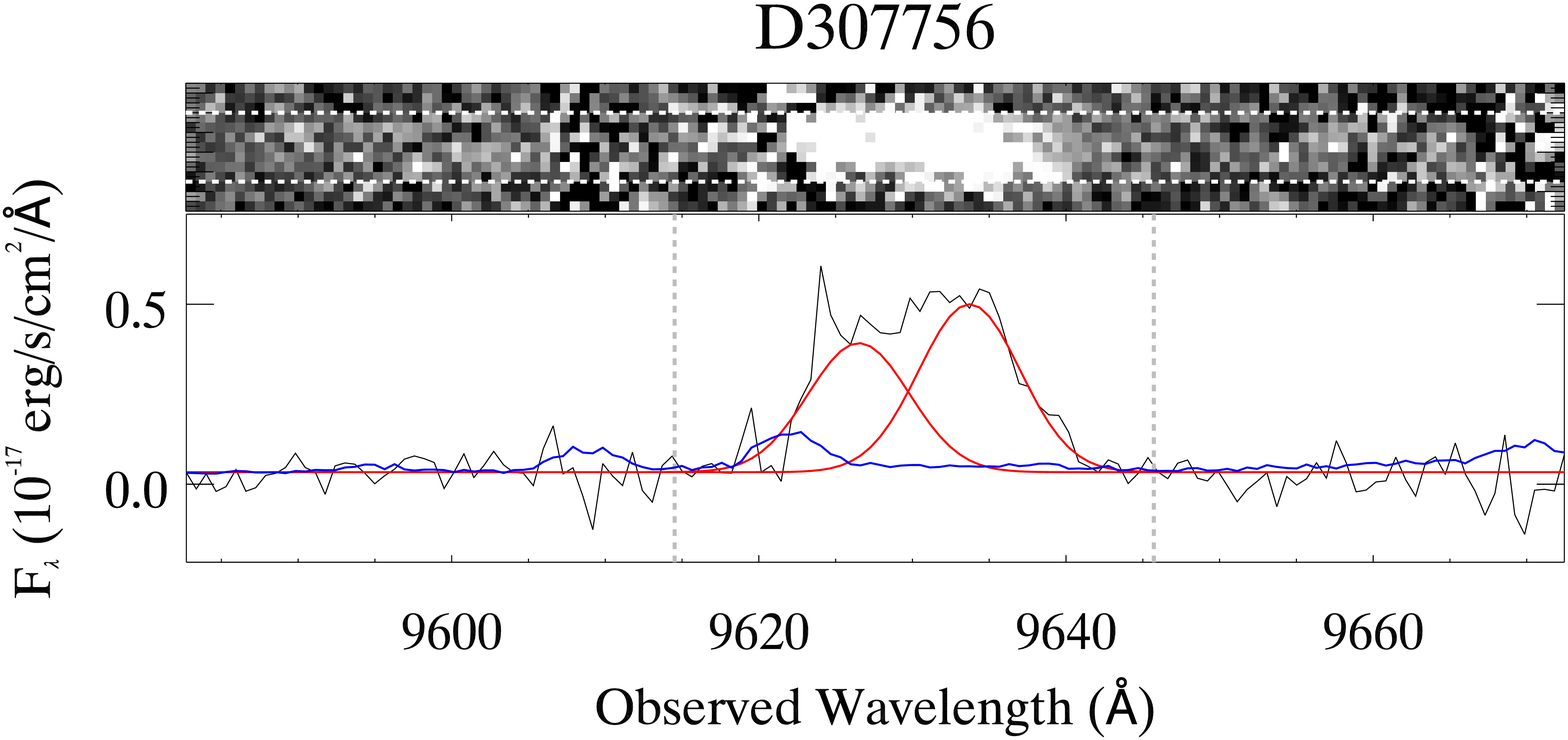}
		\includegraphics[width=0.5\textwidth,trim={0cm 0cm 0.5cm 0cm},clip]{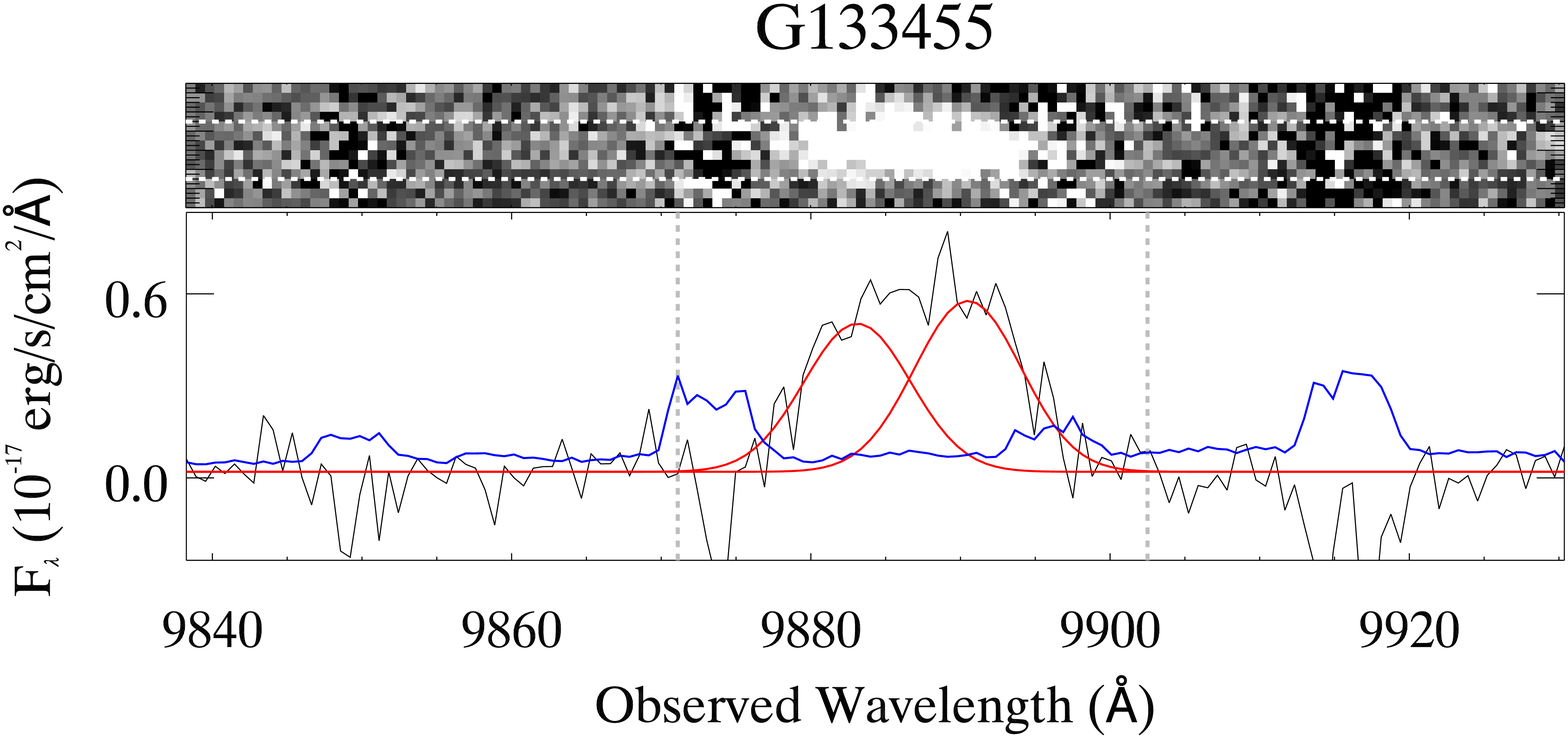}
		\includegraphics[width=0.5\textwidth,trim={0cm 0cm 0.5cm 0cm},clip]{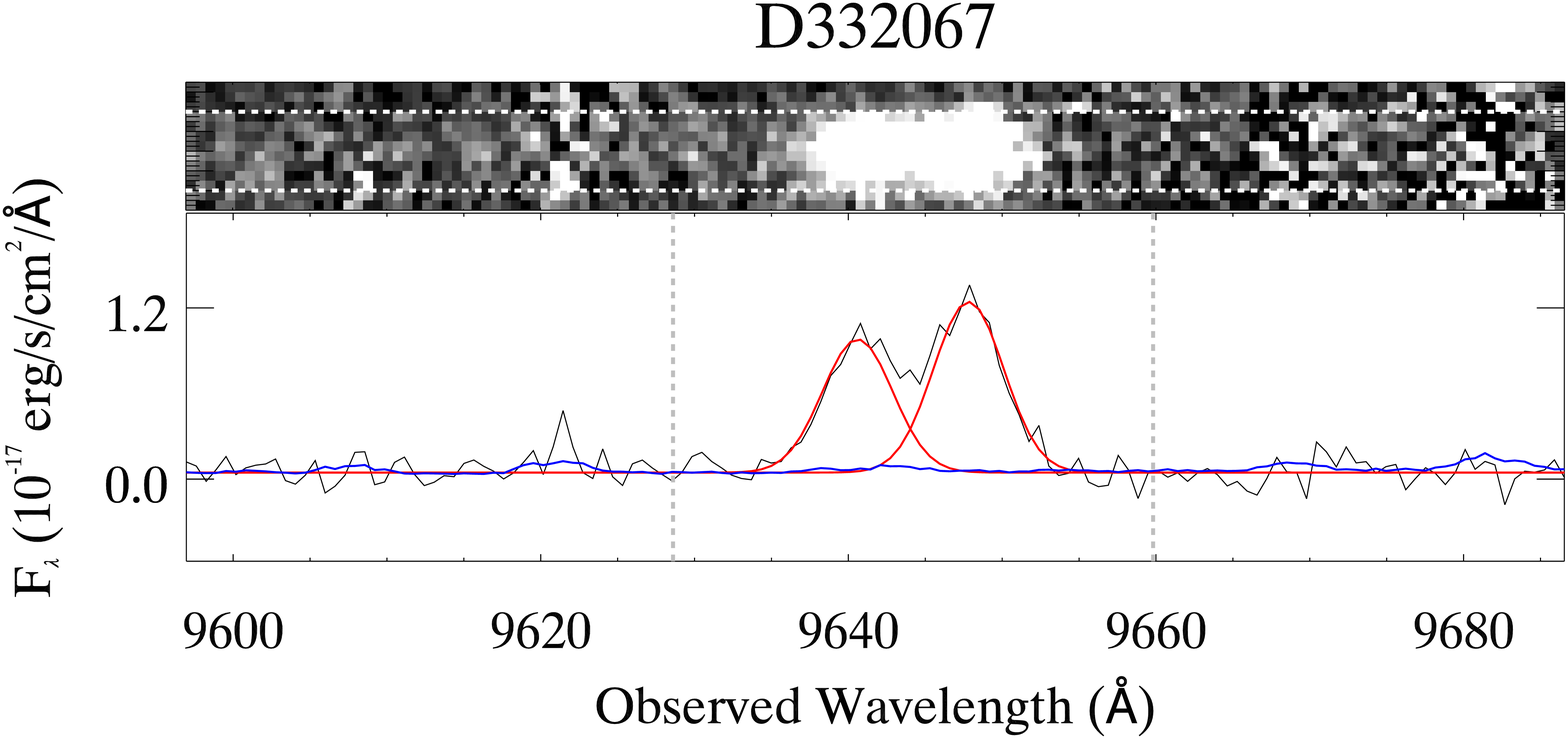}	
		\end{minipage}
	 \caption{Wavelength and flux calibrated spectra for our \OII-\Ha\ detected electron density sample. Corresponding 2D (greyscale) and 1D (black line) DEIMOS spectra are shown in the top and bottom panels respectively. Continuum and \OII emission-line fits are shown alongside the 1D spectra in red. 1D ``noise'' spectra (used to weight the fits) are indicated in blue. The effective apertures are marked for the 2D spectra (horizontal white dashed lines). The wavelength regions considered for fitting the \OII doublets are bounded by vertical grey dashed lines.}
	 \label{fig:spec}
	\end{figure*}

	\begin{figure*}
	\centering
		\begin{minipage}{\linewidth} %trim={<left> <lower> <right> <upper>}
			\includegraphics[width=0.5\textwidth,trim={0cm 0cm 0.5cm 0cm},clip]{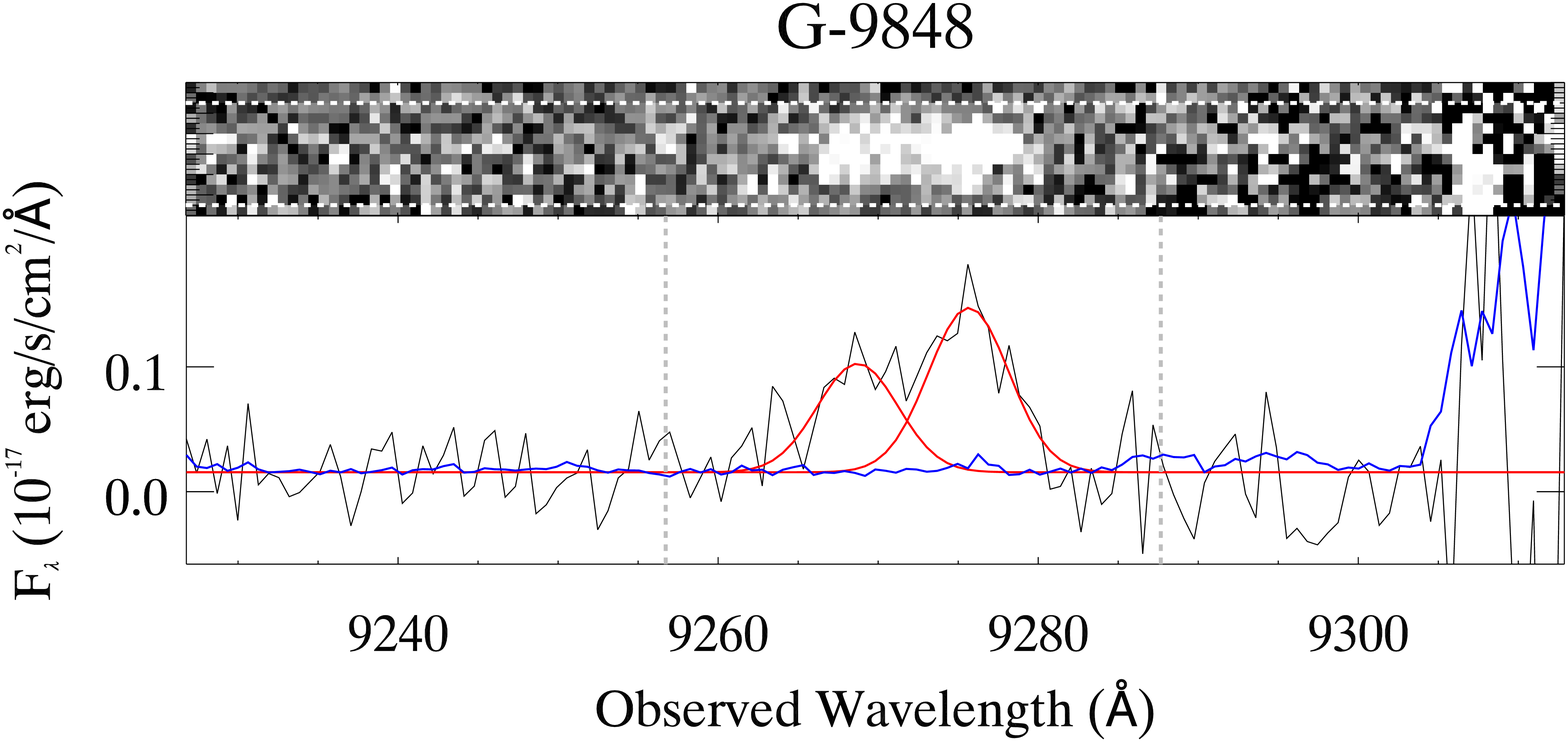}
			\includegraphics[width=0.5\textwidth,trim={0cm 0cm 0.5cm 0cm},clip]{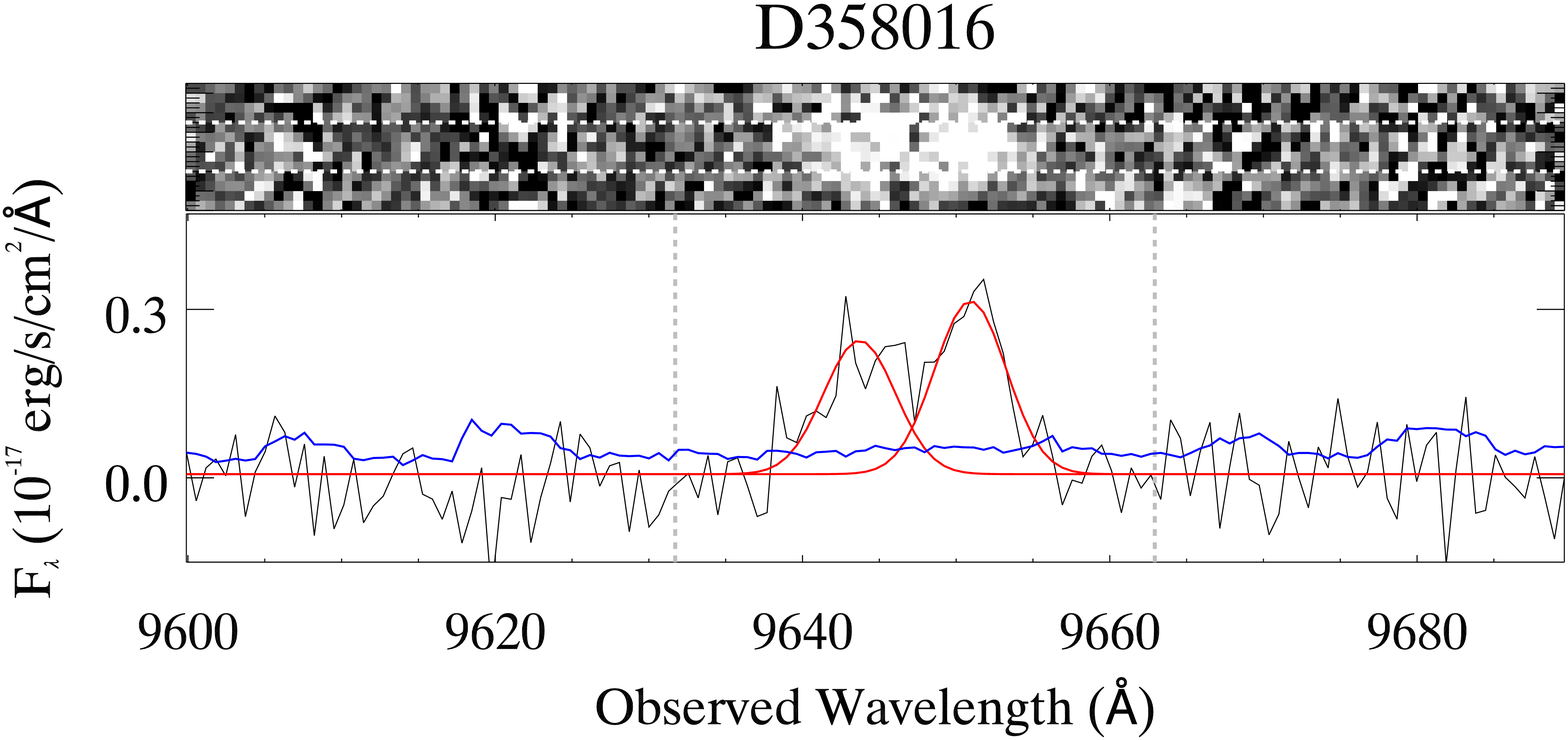}
			\includegraphics[width=0.5\textwidth,trim={0cm 0cm 0.5cm 0cm},clip]{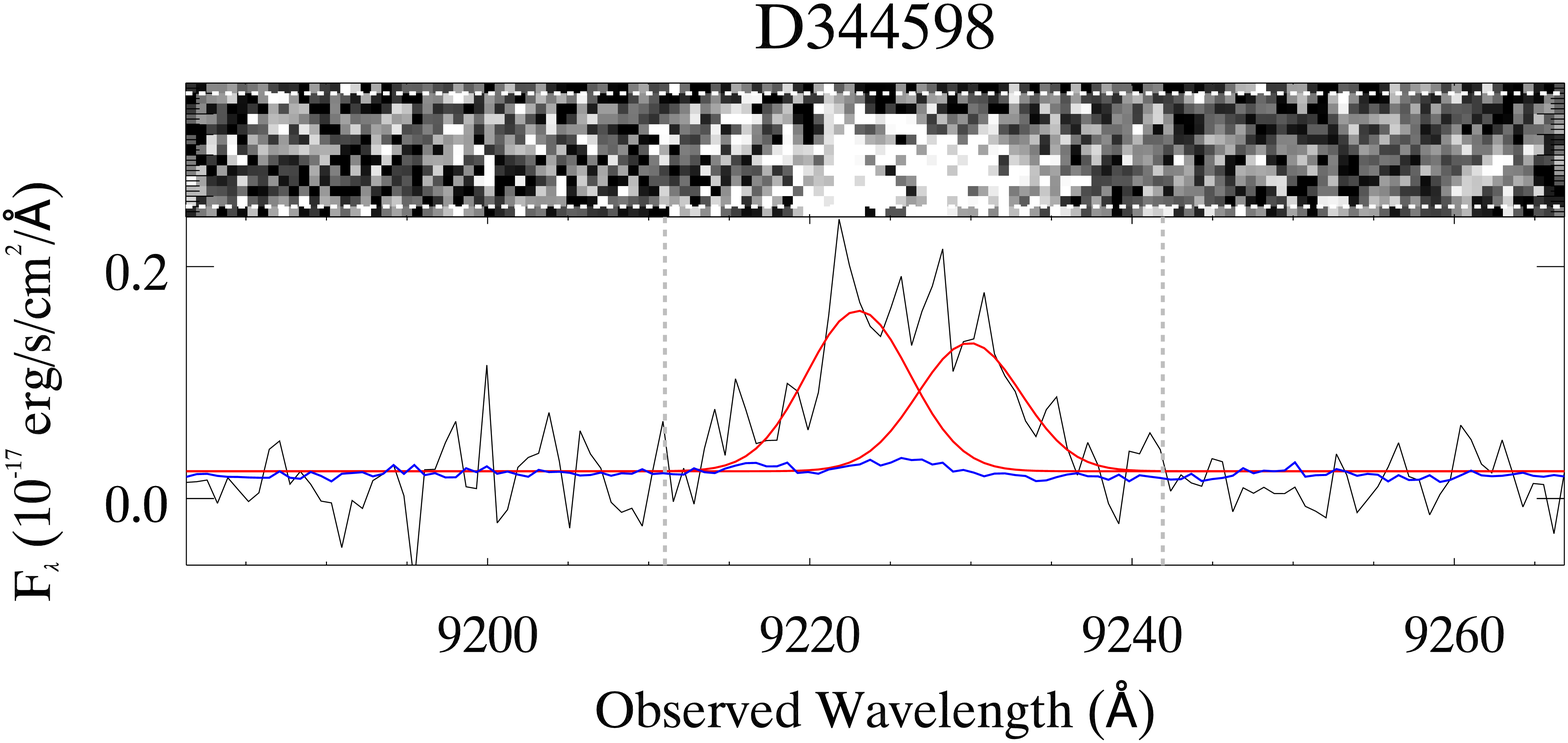}
			\includegraphics[width=0.5\textwidth,trim={0cm 0cm 0.5cm 0cm},clip]{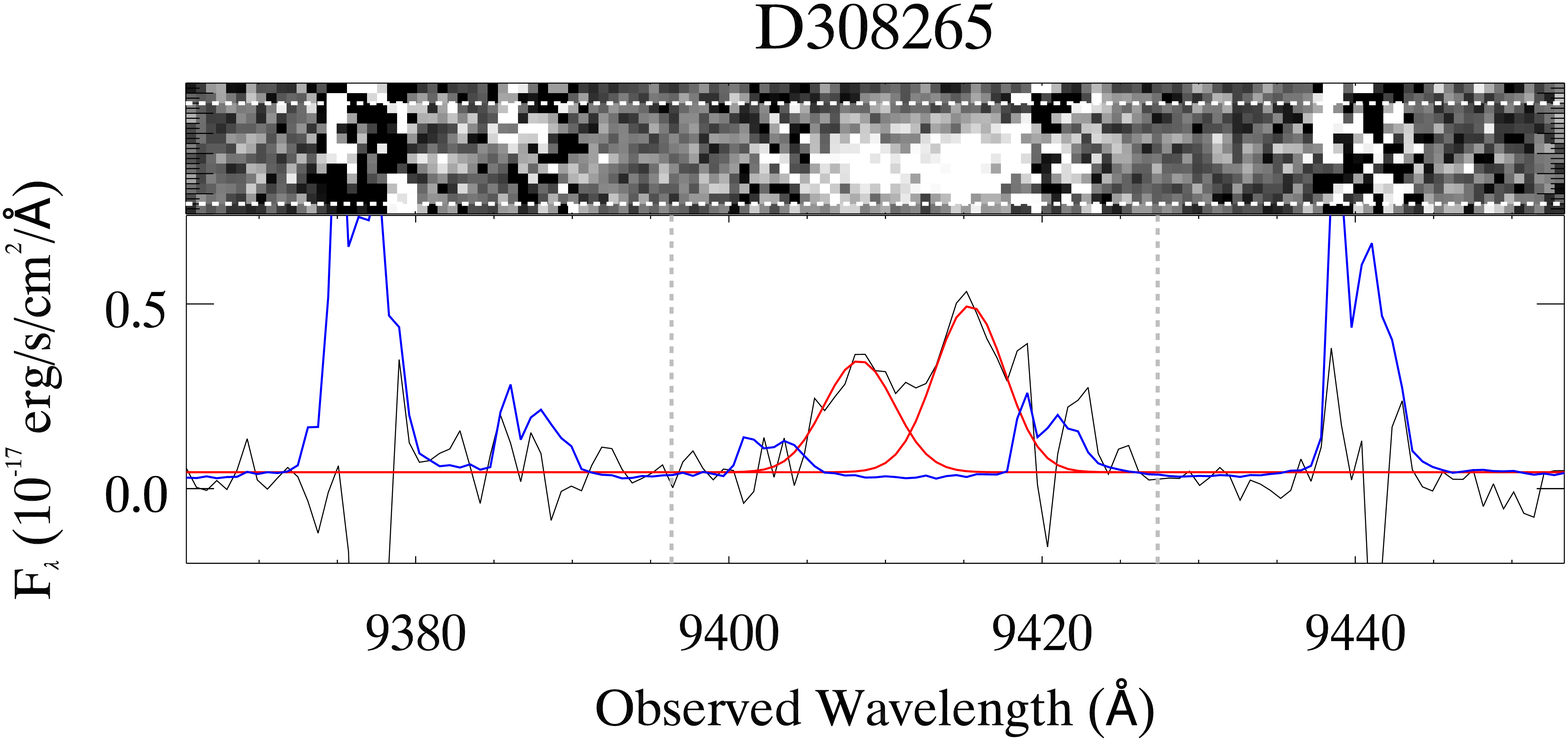}
			\includegraphics[width=0.5\textwidth,trim={0cm 0cm 0.5cm 0cm},clip]{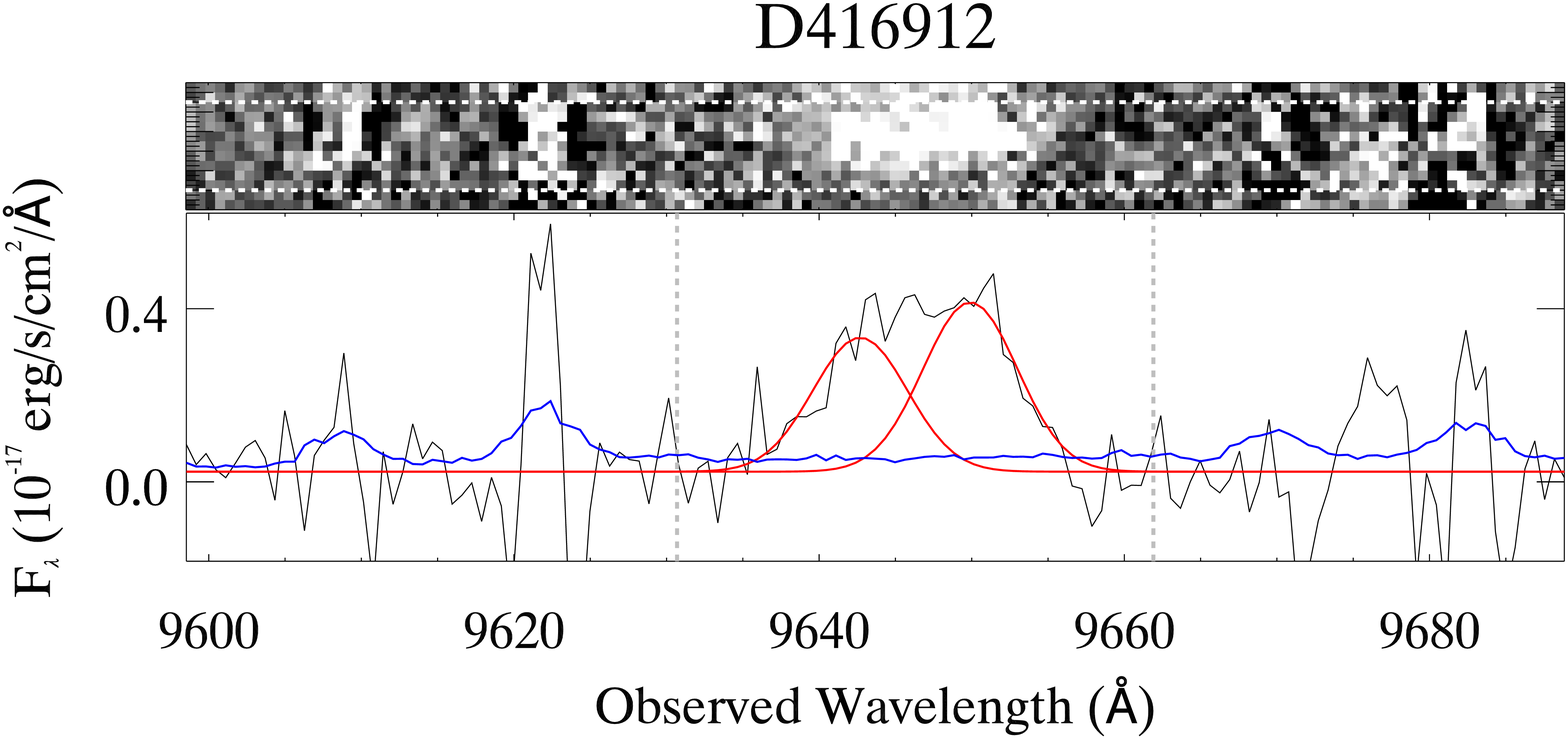}
			\includegraphics[width=0.5\textwidth,trim={0cm 0cm 0.5cm 0cm},clip]{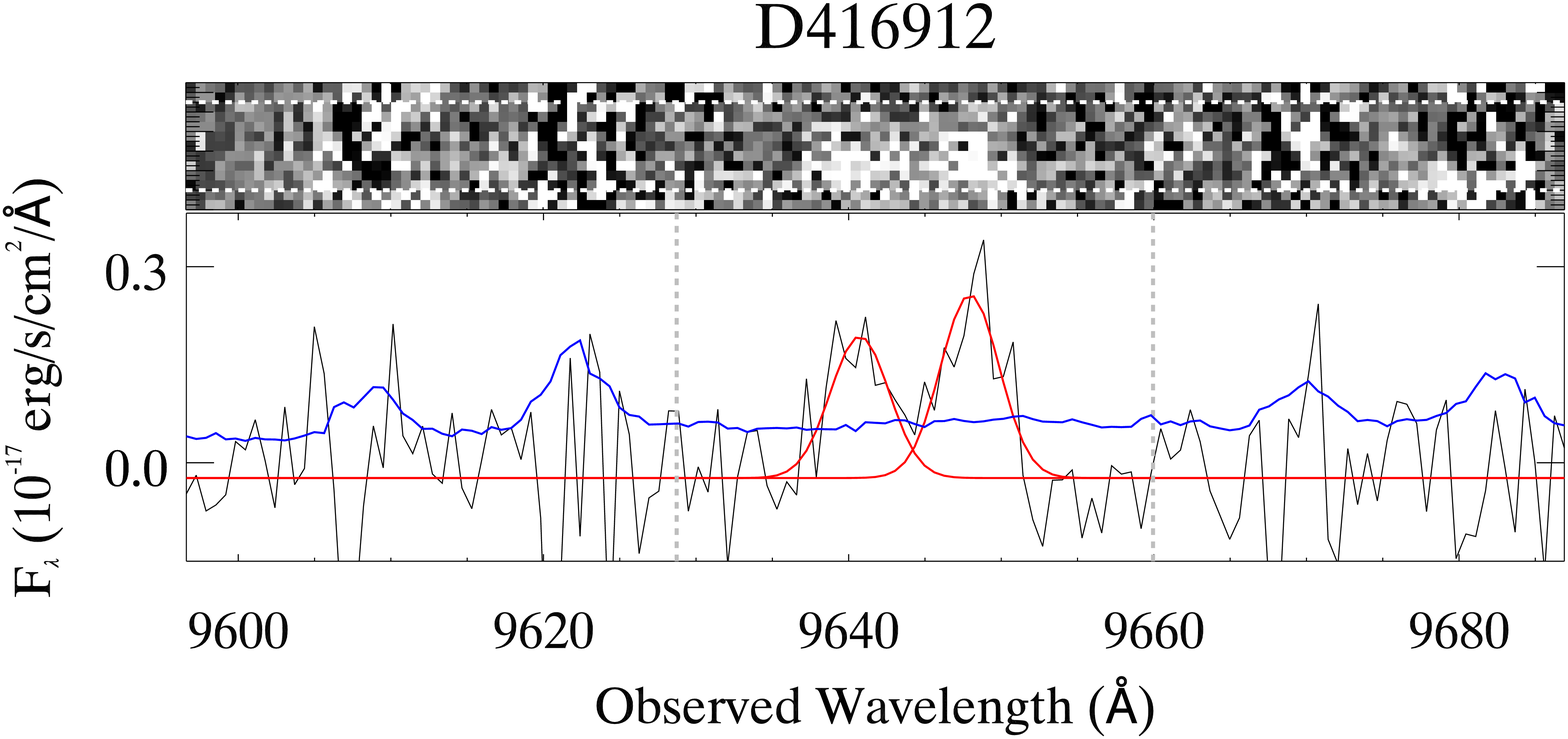}
			\includegraphics[width=0.5\textwidth,trim={0cm 0cm 0.5cm 0cm},clip]{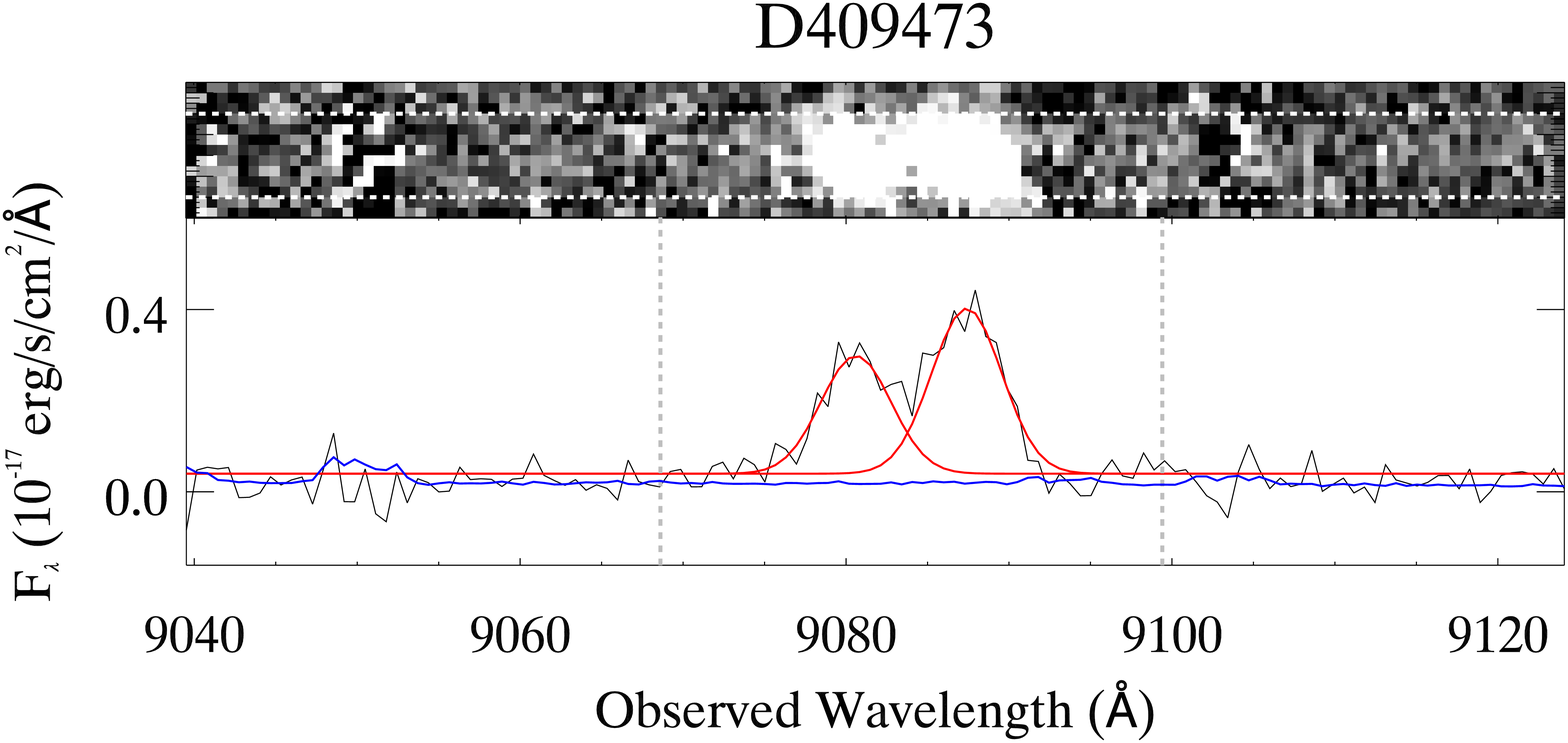}
			\includegraphics[width=0.5\textwidth,trim={0cm 0cm 0.5cm 0cm},clip]{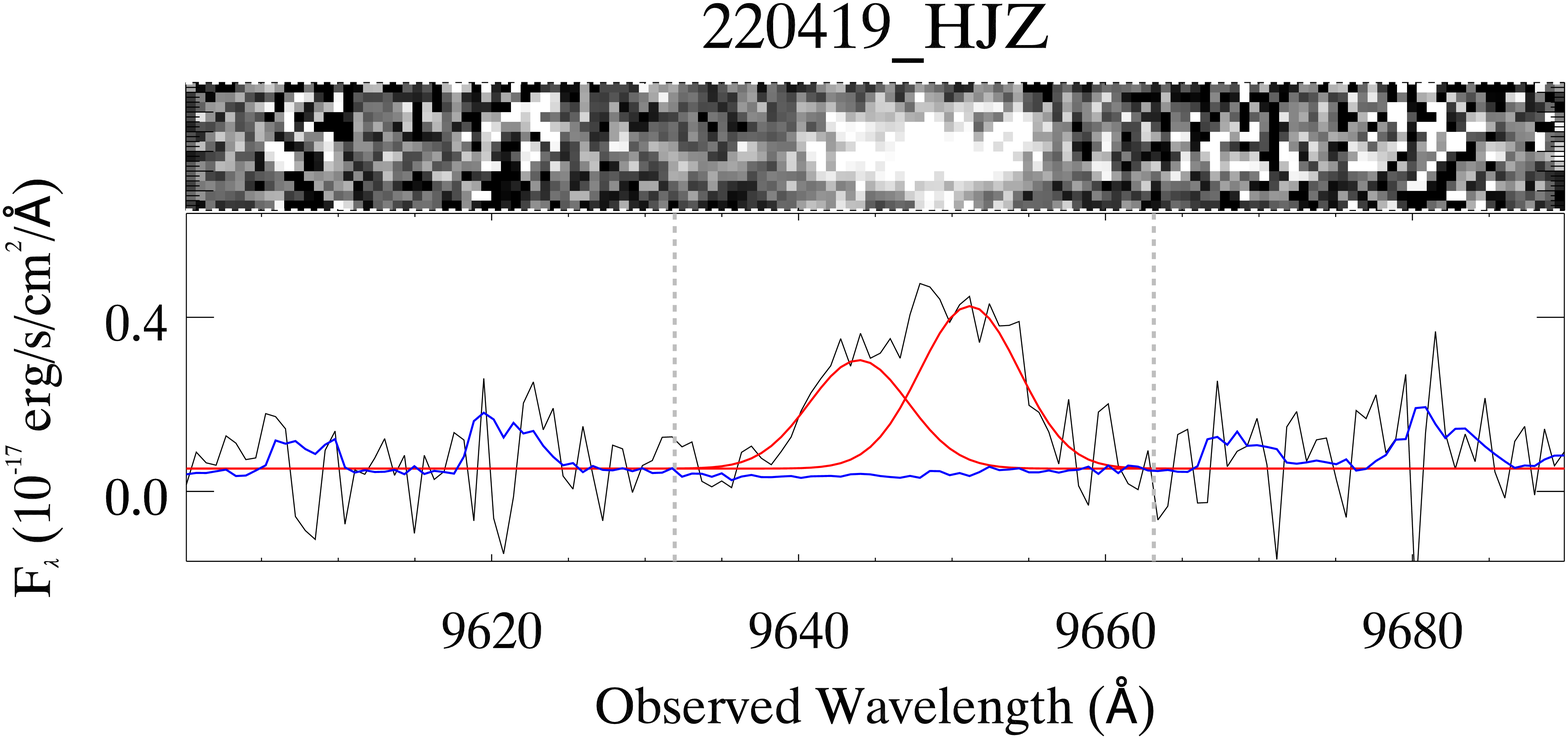}
			\includegraphics[width=0.5\textwidth,trim={0cm 0cm 0.5cm 0cm},clip]{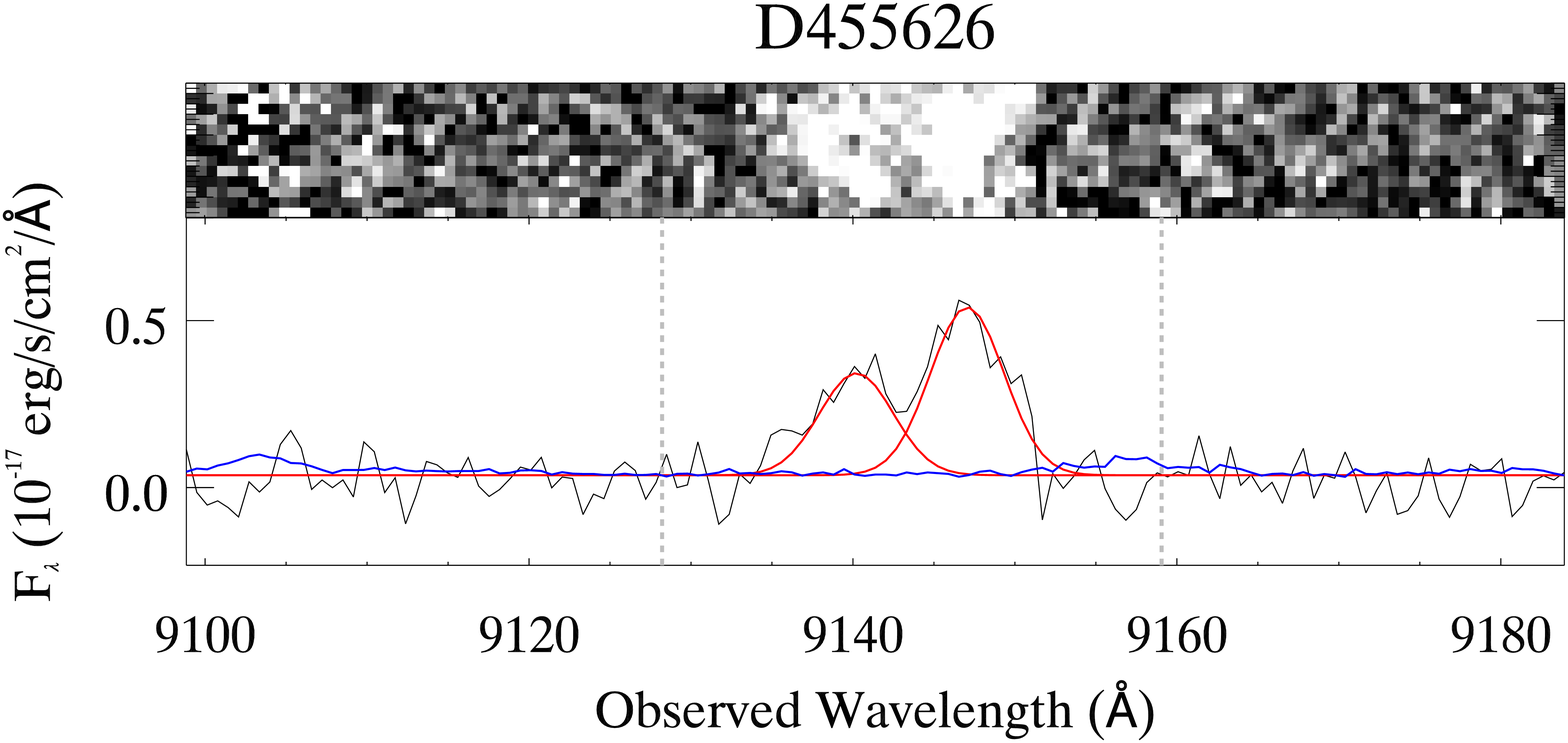}
			\includegraphics[width=0.5\textwidth,trim={0cm 0cm 0.5cm 0cm},clip]{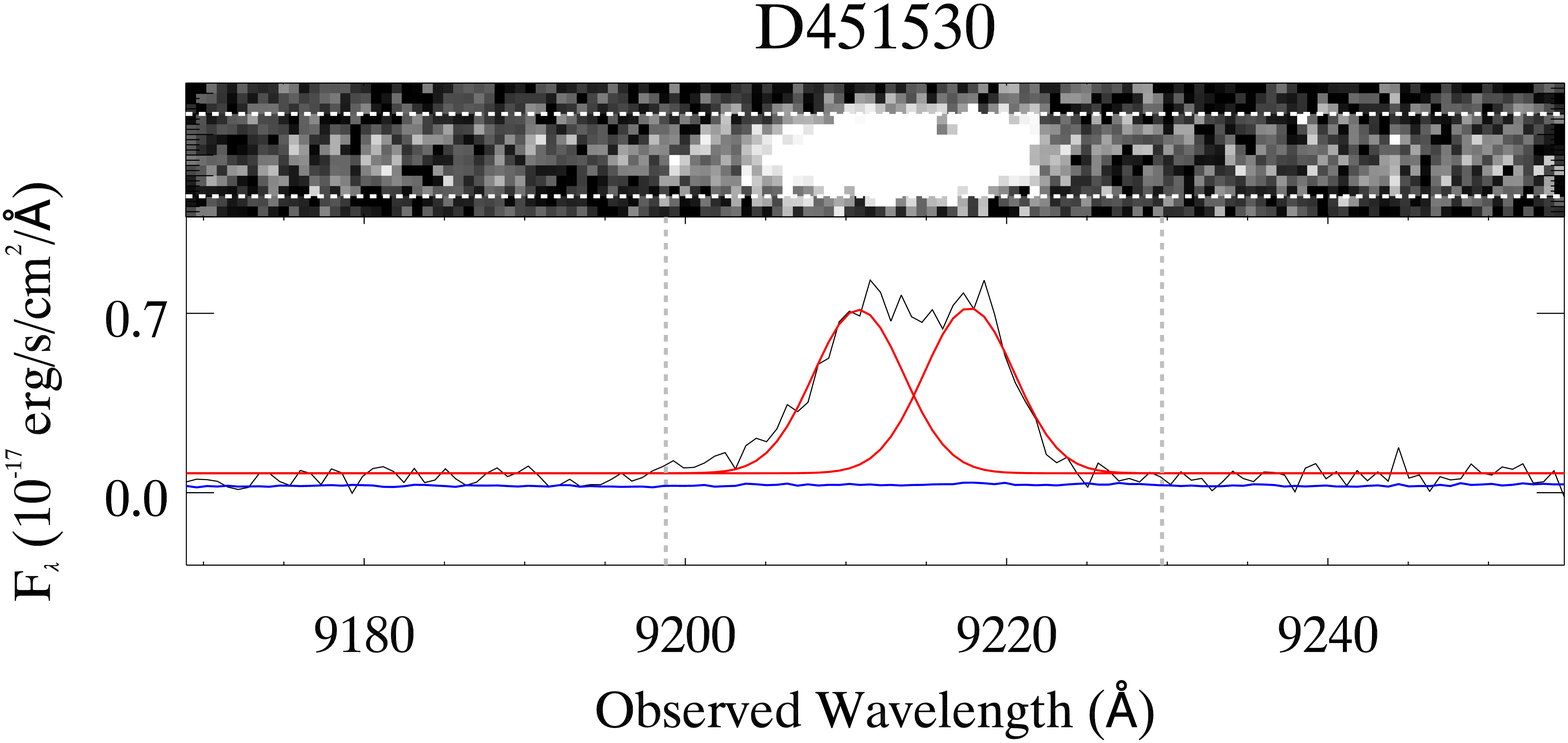}
			\end{minipage}
	 \contcaption{}
	 \label{}
	\end{figure*}

	\begin{figure*}
	\centering
		\begin{minipage}{\linewidth} %trim={<left> <lower> <right> <upper>}
			\includegraphics[width=0.5\textwidth,trim={0cm 0cm 0.5cm 0cm},clip]{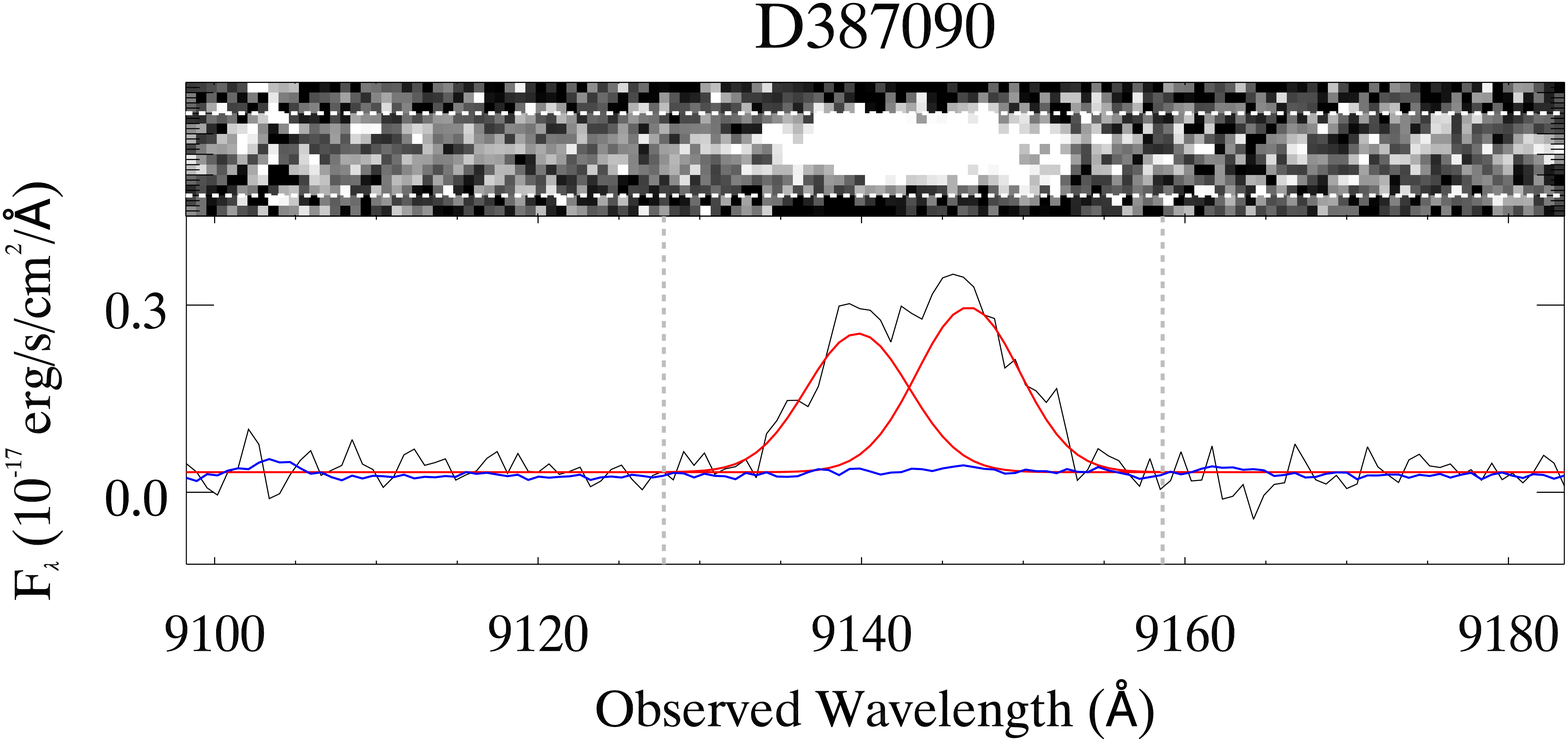}
			\includegraphics[width=0.5\textwidth,trim={0cm 0cm 0.5cm 0cm},clip]{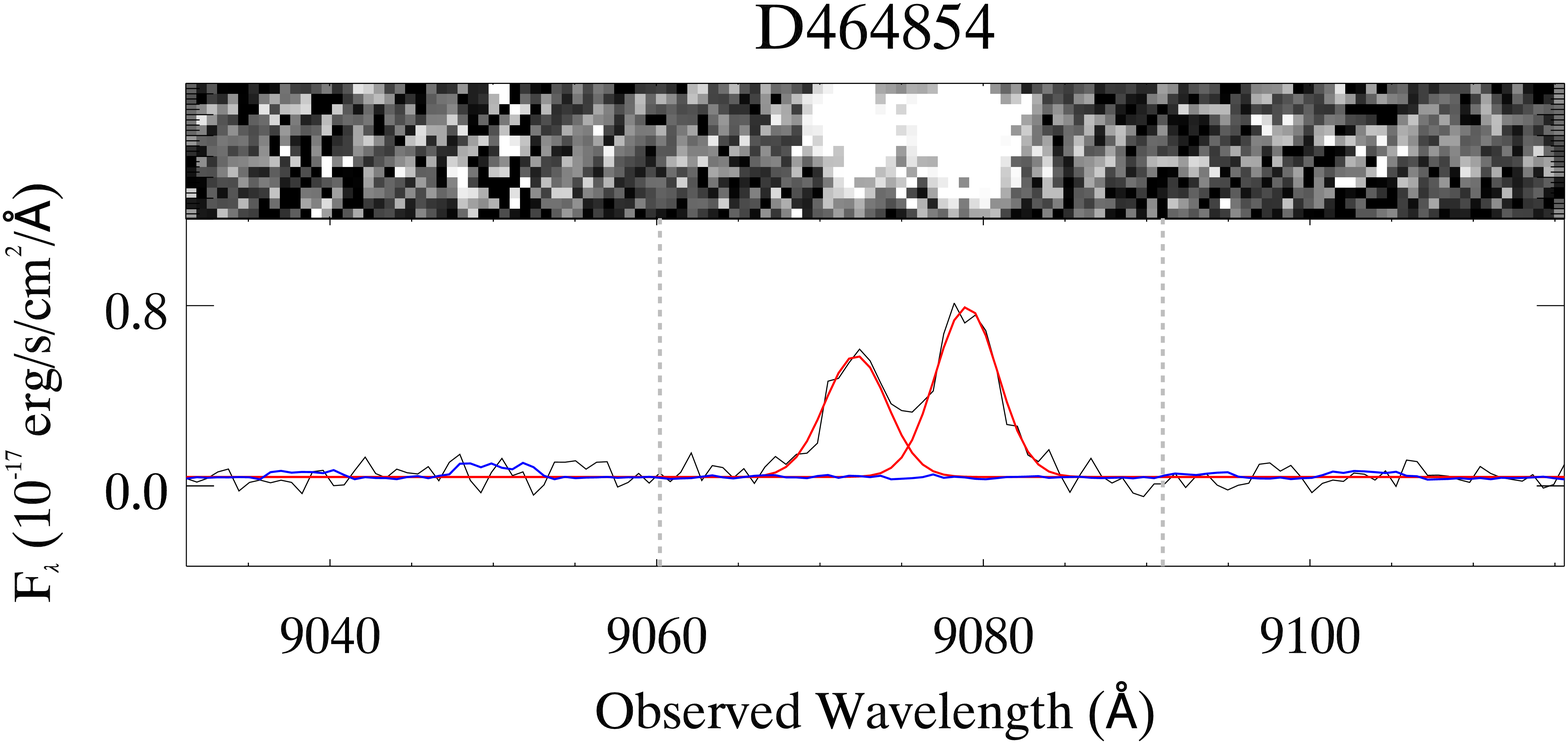}
			\end{minipage}
	 \contcaption{}
	 \label{}
	\end{figure*}

%%%%%%%%%%%%%%%%%%%%%%%%%%%%%%%%%%%%%%%%%%%%%%%%%%
{}

% Don't change these lines
\bsp	% typesetting comment
\label{lastpage}
\end{document}